\documentclass{iopart}

 \expandafter\let\csname equation*\endcsname\relax
 \expandafter\let\csname endequation*\endcsname\relax

\usepackage{graphicx,xcolor}
\usepackage{times}
\usepackage{amsmath,amssymb}
\usepackage{multirow}
\usepackage[colorlinks,citecolor=red]{hyperref}

\newcommand\braket[1]{\mathinner{\langle{\textstyle#1}\rangle}}
\newcommand{\Braket}[1]{\mathinner{\langle{\textstyle#1}\rangle}}

\newcommand{\varI}{{\mathcal{I}}}
\newcommand{\varJ}{{\mathcal{J}}}
\newcommand{\varO}{{\mathcal{O}}}
\newcommand{\tC}{\widetilde{C}}

\newcommand{\eqnref}[1]{Eq.~(\ref{#1})}
\newcommand{\eqnsref}[1]{Eqs.~(\ref{#1})}
\newcommand{\Eqnref}[1]{Equation~(\ref{#1})}
\newcommand{\Eqnsref}[1]{Equations~(\ref{#1})}

\newcommand{\figref}[1]{Fig.~\ref{#1}}
\newcommand{\figsref}[1]{Figs.~\ref{#1}}
\newcommand{\Figref}[1]{Figure~\ref{#1}}

\begin{document}

\title[Spatiotemporal Evolution of Topological Order]{Spatiotemporal Evolution of Topological Order Upon Quantum Quench Across the Critical Point}

\author{Minchul Lee$^1$, Seungju Han$^2$, and Mahn-Soo Choi$^{2,*}$}

\address{
  $^1$ Department of Applied Physics and Institute of Natural Science, College of Applied Science, Kyung Hee University, Yongin 17104, Korea \\
  $^2$ Department of Physics, Korea University, Seoul 02841, Korea
}

\ead{choims@korea.ac.kr}

\begin{abstract}
We consider a topological superconducting wire and use the string order parameter to investigate the spatiotemporal evolution of the topological order upon a quantum quench across the critical point. We also analyze the propagation of the initially localized Majorana bound states after the quench, in order to examine the connection between the topological order and the unpaired Majorana states, which has been well established at equilibrium but remains illusive in dynamical situations. It is found that after the quench the string order parameters decay over a finite time and that the decaying behavior is universal, independent of the wire length and the final value of the chemical potential (the quenching parameter). It is also found that the topological order is revived repeatedly although the amplitude gradually decreases. Further, the topological order can propagate into the region which was initially in the non-topological state. It is observed that all these behaviors are in parallel and consistent with the propagation and dispersion of the Majorana wave functions. Finally, we propose a local probing method which can measure the non-local topological order.
\end{abstract}


\maketitle

\section{Introduction}
\label{sec:introduction}

Traditionally, Landau's symmetry-breaking theory
\cite{Landau37a,Landau37b,Landau80a} has long provided a universal paradigm
for the states of matter and their transitions. In this paradigm, continuous
phase transitions are driven by the spontaneous symmetry breaking and
naturally described by local order parameters.
In recent decades, condensed-matter physics has witnessed the breakdown of the symmetry-breaking theory in describing symmetry-protected topological states.
One of the most common and earliest examples is the quantum Hall effect \cite{Klitzing80a,Tsui82a}, where different quantum Hall states have the same symmetry.
Topological phase transitions involve the change in internal topology rather than symmetry breaking \cite{Nayak08a,Hasan10a}. Necessarily, topological states are classified by topological quantum numbers.
For example, different quantum Hall states are classified by the topological Chern number \cite{Laughlin81a,Laughlin83b}, and topological insulators and superconductors are characterized by the number of gapless boundary (surface, edge or endpoint) states \cite{Kane05a,Kane05b,Hasan10a,Qi11a} separated from gapped bulk states.

From the dynamical point of view, the classification in terms of
\emph{discrete} topological quantum numbers puts a serious limitation. For
example, let us ask the question, how long does it take for a topological order
to form? Namely, how does the topological order emerge or disappear temporally
when system parameters are quenched across the critical point? Within Landau's
paradigm, the approach to the corresponding questions is conceptually clear
because the local order parameters take continuous values and their dynamics
are governed by a differential equation of motion, so called the time-dependent
Ginzburg-Landau equation \cite{Stephen64a,Abrahams66a,Hohenberg77a}.

Another conceptual difficulty in the dynamical description of topological order arises from the fact that the topological quantum numbers or similar classifications concern about the ground state(s). In order to describe the full aspects of the temporal evolution of the topological order, one has to take into account the excited states as well as the ground states.

To overcome these issues in the dynamical theory of topological order, a few different approaches have been carried out previously:
In Ref.~\cite{Perfetto13a}, the fidelity of the final state with respect to the
initial state has been examined. Similarly, the survival and revival
probabilities \cite{Rajak2014apr,Hegde15a} or the Loschmidt echo
\cite{Vasseur14a} of the Majorana states after the quench have been also
investigated.
In particular, the sudden coupling of the localized Majorana state to the
normal metallic (gapless) lead gives rise to universal features in the evolution
of the Majorana wave function \cite{Vasseur14a}.
For a ladder system studied in Ref.~\cite{DeGottardi11a}, the
number of excited vortices in plaquettes was inspected.
It has also been suggested to use the Kibble-Zurek mechanism, which was originally put forward to study the cosmological phase transitions of the early Universe \cite{Kibble76a,Kibble80a} and later extended to classical phase transitions \cite{Zurek85a,Zurek96a}. This approach is particularly simple and insightful as it is only based on the competition between the internal and driving time scales. It has been shown that the original Kibble-Zurek scaling form does not hold for the topological phase transitions \cite{Bermudez09a,Bermudez10a} but one can generalize it by properly taking into account the multi-level structure due to the Majorana bound states \cite{ChoiMS15b}.

However, none of these approaches directly measures the dynamic topological order; they either resort to the fidelity to the ground state or to the number of topological defects. Here we note that in one spatial dimension one can draw a close analogy with the conventional order by introducing a continuous parameter for the topological order~\cite{Endres11a}.
The cost is that unlike the conventional order parameter the so-called \emph{string order parameter} for the topological order is \emph{nonlocal}: It is defined as the the expectation value of a product of consecutive Majorana operators in a certain range with respect to the dynamical wave function.
It naturally captures the dynamical evolution of the topological order reflected in the wave function, and is considered to be more suitable for experimental observations~\cite{Endres11a}.

On these grounds, in this work we adopt the string order parameter \cite{Haegeman12a,Pollmann12a,Bahri2014apr} to investigate the spatiotemporal evolution of the topological order upon a quantum quench across the critical point in a topological superconducting wire~\cite{Kitaev2001oct,Alicea10a,Alicea11a,Alicea12a}.
In addition, we separately analyze the propagation of the initially localized Majorana bound states after quenching.
By comparing the propagations of the topological order and the Majorana states, we examine the connection between them in the dynamical situations.
It is stressed that while the close connection between the topological order and the Majorana bound states has been well established for the \emph{ground state}, i.e., at equilibrium~\cite{Kitaev2001oct,Alicea10a,Alicea11a,Alicea12a}, it is not obvious at all in the dynamical situations, which involve \emph{excited states}.

We find that after the
quench toward the nontopological phase, the string order parameters decay over a finite time before
they vanish completely.
Notably, this early decaying behavior is universal in the sense that it does not depend on the wire length and the final value of the chemical potential (the quenched parameter).
We also find that the topological order is revived repeatedly although the amplitude gradually decreases. More interestingly, the topological order propagates into the regions which were initially prepared in the non-topological state.
It is observed that all these behaviors of the dynamical topological order
are in parallel and consistent with the propagation and dispersion of the Majorana wave functions.
Finally, we propose a local probing method which can measure the topological order which is nonlocal in nature.

The rest of the paper is organized as following: Section~\ref{sec:tsw}
describes the model Hamiltonian for the topological superconducting wire and
defines the string order parameters. For later use, it summarizes the
mathematical and physical properties of the string order
parameters. Section~\ref{sec:uniform} studies the time evolution of the string
order parameter in a uniform wire. It discusses the dynamical aspects of the
topological order in terms of the propagation of the Majorana wave
functions. Section~\ref{sec:propagation} investigate the case where only the
central part of the superconducting wire is initially in the topological phase
and the outer part is topologically trivial so that the propagation of
  the topological order is examined. Section~\ref{paper::sec:5} proposes
possible experimental methods to observe our findings. Finally,
Section~\ref{sec:conclusion} concludes the paper.

\section{Topological Superconducting Wires and Topological Order}
\label{sec:tsw}

\subsection{Model}
\label{sec:model}

\begin{figure}[!b]
  \centering
  \includegraphics[width=.8\textwidth]{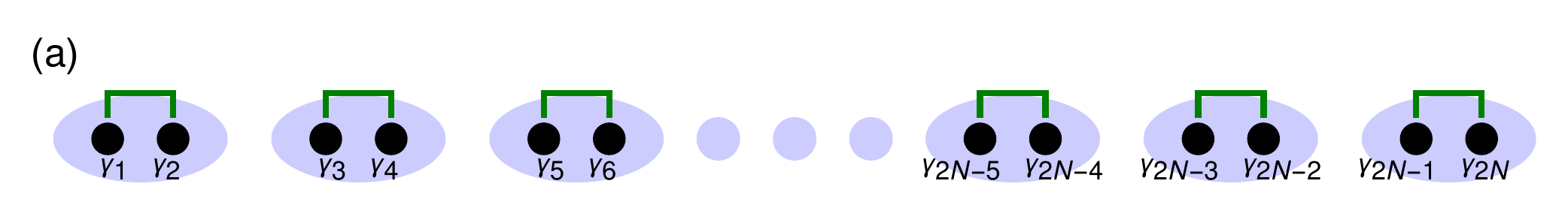}\\
  \includegraphics[width=.8\textwidth]{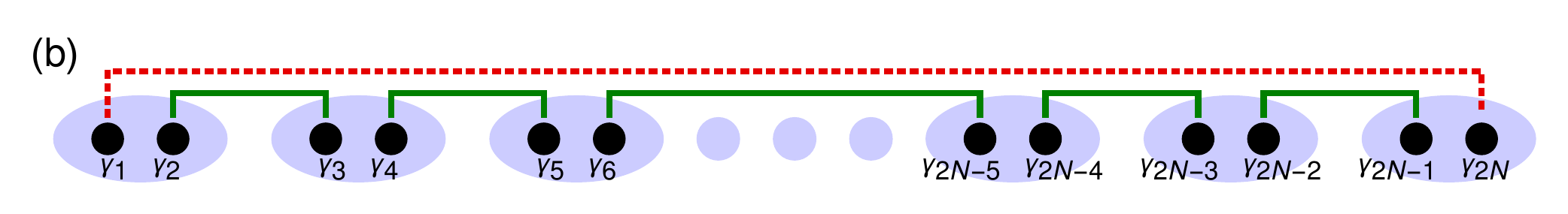}\\
  \includegraphics[width=.8\textwidth]{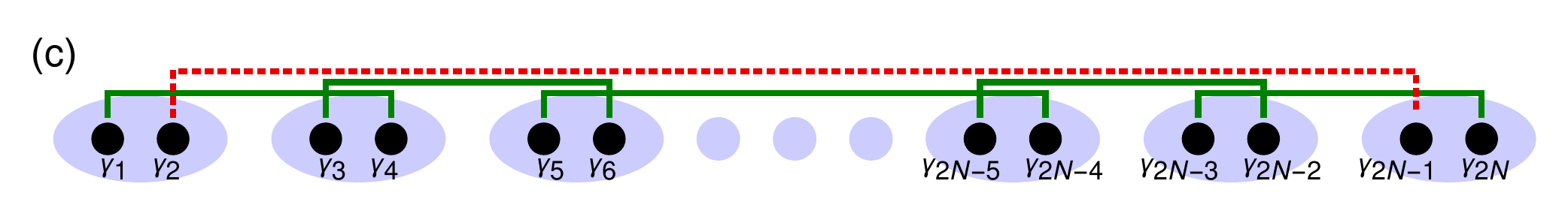}
  \caption{Majorana chain representation of one-dimensional topological
    superconducting wire. Each pair of Majorana fermions, $\gamma_{2j-1}$ and
    $\gamma_{2j}$ in a circle forms the site fermion $c_j$. The lines
    connecting Majorana fermions represent the coupling between them in three
    special conditions: (a) $w = \Delta = 0$ and $\mu\ne0$, (b) $\Delta = w$
    and $\mu = 0$, and (c) $\Delta = -w$ and $\mu = 0$. The non-local
    correlations (red dotted lines) between otherwise isolated Majorana
    fermions arise only when the fermion parity condition is imposed.}
  \label{fig:Majoranachain}
\end{figure}

In order to explore the time evolution of topological order under a quench, we
consider Kitaev's spinless $p$-wave topological superconducting wire with
finite number of sites $N$ and with open ends described by the tight-binding
Hamiltonian \cite{Kitaev2001oct}
\begin{subequations}
  \begin{align}
    H(t)
    & =
    \sum_{j=1}^{N-1}
    \frac12 \left(\Delta c_j c_{j+1} - w c_j^\dag c_{j+1} + (h.c.)\right)
    - \sum_{j=1}^{N} \mu_j(t) \left(c_j^\dag c_j - \frac12\right)
    \label{eq:model}
    \\
    & =
    \frac{i}{4}
    \sum_{j=1}^{N-1}
    \left[
      (\Delta + w) \gamma_{2j} \gamma_{2j+1}
      + (\Delta - w) \gamma_{2j-1} \gamma_{2j+2}
    \right]
    - \frac{i}{2} \sum_{j=1}^N \mu_j(t) \gamma_{2j-1} \gamma_{2j}
  \label{eq:model:Majorana}
  \end{align}
\end{subequations}
with the hopping amplitude $w>0$, the superconducting gap $\Delta$, and the
site-dependent chemical potential $\mu_j(t)$. The phase of the
superconducting order parameter is neglected since it can be always gauged
away. The fermion operator $c_j$ on site $j$ can be decomposed into the
superposition of two Majorana fermion operators: $c_j = (\gamma_{2j-1} + i
\gamma_{2j})/2$, satisfying $\gamma_k = \gamma_k^\dag$ and
$\{\gamma_k,\gamma_{k'}\} = 2\delta_{kk'}$. This system preserves the fermion
parity, or $\mathbb{Z}_2$ symmetry, defined by
\begin{equation}
  \label{eq:parity}
  P
  = (-1)^{\sum_{j=1}^N c_j^\dag c_j}
  = \prod_{j=1}^N (- i \gamma_{2j-1} \gamma_{2j})
\end{equation}
since all the terms in \eqnref{eq:model} either preserve the charge number or change
it by even numbers.  This system has a topological invariant, known as the
winding number, which takes values $\pm 1$ and 0. First consider the uniform
and static case with $\mu_j(t) = \mu$. For $|\mu| > w$, the fermion parity is
not broken and the system is in non-topological phase with no unpaired Majorana
fermions as demonstrated in \figref{fig:Majoranachain}(a). On the other hand,
for $|\mu| < w$ and $\Delta\ne0$, there are topological phases with broken
$\mathbb{Z}_2$ symmetry, and the topological invariant is 1 for $\Delta > 0$ or
$-1$ for $\Delta < 0$. In the topological phases two unpaired Majorana modes
arise at two ends of open and long chains, as shown
\figsref{fig:Majoranachain}(b) and (c): For example, for $\Delta = w$ and $\mu
= 0$, the Majorana fermions operators $\gamma_1$ and $\gamma_{2N}$ are absent
in the Hamiltonian (\ref{eq:model:Majorana}), giving two zero-energy
modes. Similarly, for $\Delta = -w$ and $\mu = 0$, $\gamma_2$ and
$\gamma_{2N-1}$ are isolated. In these special conditions or for infinitely
long chain, the ground state is doubly degenerate with definite fermion parity
$P = \pm1$. The two-fold degeneracy coming from $\mathbb{Z}_2$ symmetry cannot
be lifted by small local perturbations unless they involve odd number of
fermion operators such as $c_j$ or $c_j c_k^\dag c_i$, which is the reason why
the phase is called topological. The phase transition between topological and
non-topological phases occurs at $|\mu| = w$, where the system becomes gapless.

\subsection{Topological Order}

In our study we adopt the nonlocal string order parameter proposed by Bahri and
Vishwanath \cite{Haegeman12a,Pollmann12a,Bahri2014apr} to measure the topological order in quench
dynamics. The physical meaning of the string order parameter, especially in the
Kitaev model (\ref{eq:model:Majorana}), can be understood in its
counterpart spin-1/2 model. Via the Jordan-Wigner transformation, the Kitaev
model is mapped onto the spin XY model in a transverse magnetic field described
by
\begin{equation}
  H
  =
  - \sum_{j=1}^{N-1}
  \left(J_x \sigma_j^x \sigma_{j+1}^x + J_y \sigma_j^y \sigma_{j+1}^y\right)
  - h_z \sum_{j=1}^N \sigma_j^z
\end{equation}
with $J_x = (w + \Delta)/4>0$, $J_y = (w - \Delta)/4>0$, and $h_z =
-\mu/2$. The $\mathbb{Z}_2$ symmetry in spin model corresponds to a global spin
flip in the $\sigma^x$ basis: $P = \prod_{j=1}^N \sigma_j^z$.

\begin{figure}[!t]
  \centering
  \includegraphics[width=.5\textwidth]{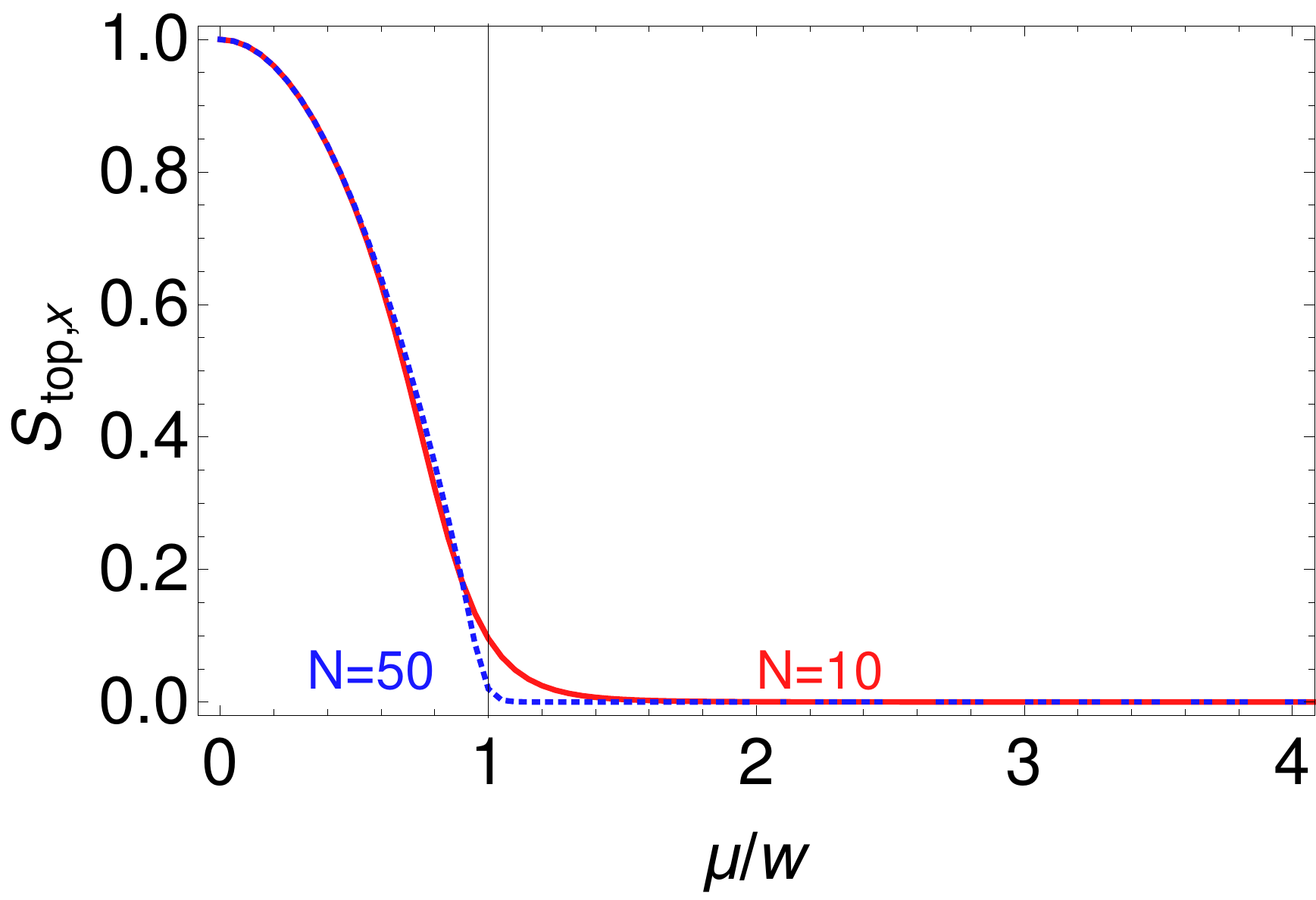}
  \caption{String order parameter $S_{\mathrm{top},x}$ for $N=10$ (red solid line)
    and $50$ (blue dotted line) topological superconducting wires as a function
    of the chemical potential $\mu$ in equilibrium at zero temperature. Here we
    set $w = \Delta = 1$. The order parameter exhibits a second-order-type
    transition behavior: it is finite in the topological phase ($\mu < w$) and
    vanishes in the non-topological phase ($\mu > w$).}
  \label{fig:sop:eq}
\end{figure}

Consider the Ising case with $w = \Delta$ (or $J_y = 0$) for simplicity. This
system, like its counterpart (\ref{eq:model}), experiences a phase transition
at $J_x = |h_z|$ (that is, $|\mu| = w$) from the spin ordered phase ($J_x >
|h_z|$) to the disordered phase ($J_x < |h_z|$). In the spin ordered phase or
ferromagnetic phase, the $\mathbb{Z}_2$ symmetry is broken like the topological phase in its counterpart. The spin order in the ferromagnetic phase
is reflected in the spin-$x$ correlation between two spins at end points:
for $N\gg 1$ \cite{Pfeuty1970mar},
\begin{equation}
  \Braket{\sigma_1^x \sigma_N^x}
  =
  \begin{cases}
    \displaystyle
    (1 - J_x/|h_z|)^{\frac14} + \varO(1/N), & J_x > |h_z|
    \quad\text{(spin ordered phase)}
    \\
    \varO(1/N), & J_x < |h_z|
    \quad\text{(spin disordered phase)}.
  \end{cases}
\end{equation}
So the non-vanishing value of the spin correlation proves the existence of the
ferromagnetic order over the whole chain. Noting that the spin order in the
spin model is no other than the topological order in the original model, the
spin correlation can be used to measure the topological order. Indeed, via the
Jordan-Wigner transformation, the spin correlation can be expressed in terms of
the Majorana operators as
\begin{equation}
  \label{eq:sop:x}
  S_{\mathrm{top},x}
  =
  \braket{(-i\gamma_2)
    \left[\prod_{j=2}^{N-1} (-i\gamma_{2j-1} \gamma_{2j})\right]
    \gamma_{2N-1}}
  = \braket{\prod_{j=1}^{N-1} (-i \gamma_{2j} \gamma_{2j+1})}.
\end{equation}
This is the string order parameter for the topological order. Note that the
order parameter is based on the nonlocal operator, a string of Majorana
operators, reflecting the fact that the
topological order is nonlocal property. On the contrary, its counterpart in the spin model, $\Braket{\sigma_1^x
  \sigma_N^x}$, is defined by local operators. The subscript $x$ comes from the fact
that the string order parameter (\ref{eq:sop:x}) comes from the spin-$x$
correlation. Clearly, this string order parameter becomes exactly
one for the configuration in \figref{fig:Majoranachain}(b) since the Majorana
fermions between neighboring sites, $\gamma_{2j}$ and $\gamma_{2j+1}$, are
maximally paired. \Figref{fig:sop:eq} demonstrates that the string order
parameter in the Ising case displays the second-order transition behavior.

In general, the ferromagnetic ordering can arise along the spin-$y$ direction
if the spin XY model ($J_x,J_y \ne 0$) is taken into account or when
the system evolves with time under the transverse magnetic field so that the
spins precess. Specifically, the spin-$y$ correlation between two edge spins,
$\Braket{\sigma_1^y \sigma_N^y}$, is
written via the Jordan-Wigner transformation as
\begin{equation}
  \label{eq:sop:y}
  S_{\mathrm{top},y}
  =
  \braket{\gamma_1
    \left[\prod_{j=2}^{N-1} (-i\gamma_{2j-1} \gamma_{2j})\right]
    (i\gamma_{2N})}
  = (-1)^{N-1} \braket{\prod_{j=1}^{N-1} (-i\gamma_{2j-1} \gamma_{2j+2})}.
\end{equation}
Interestingly, this string order parameter vanishes for the configuration in
\figref{fig:Majoranachain}(b) but becomes one for that in
\figref{fig:Majoranachain}(c). Therefore, the spin-$x$ and -$y$ correlations
capture two different kinds of topological order for which the topological
invariant is 1 or $-1$, respectively. Hence, the sum of two order parameters
defines the total topological order:
\begin{equation}
  \label{eq:sop}
  S_\mathrm{top}
  = S_{\mathrm{top},x} + S_{\mathrm{top},y}
\end{equation}
which is no other than the spin correlation $\Braket{\sigma_1^x \sigma_N^x +
  \sigma_1^y \sigma_N^y}$ in the $xy$ plane perpendicular to the transverse
field.

From the dynamical point of view, the most fascinating aspect of this string order parameter is that it is
defined with respect to the wave function, not to the Hamiltonian. The
so-called topological invariants which are commonly used to clarify the
topological property of the system \emph{at equilibrium} are basically the properties of
the Hamiltonian.  Therefore they are inadequate for the study of the evolution
of the topological order triggered by the change of the Hamiltonian with time
such as quench. Quite contrary, the string order parameter $S_\mathrm{top}$ is the
expectation value of Majorana operators with respect to the wave function which
evolves with time, so it can capture the dynamical evolution of the topological
order in the wave function.

The string order parameters can be defined over a part of the chain as well as the whole. For
later use, we define the string order parameters for the region from site $j_1$
to $j_2$ as
\begin{subequations}
  \begin{align}
    \label{eq:sop:x:region}
    S_{\mathrm{top},x}(j_1,j_2)
    & = \braket{\prod_{j=j_1}^{j_2-1} (-i \gamma_{2j} \gamma_{2j+1})}
    \\
    S_{\mathrm{top},y}(j_1,j_2)
    & = (-1)^{j_2-j_1} \braket{\prod_{j=j_1}^{j_2-1} (-i\gamma_{2j-1} \gamma_{2j+2})}
    \\
    S_\mathrm{top}(j_1,j_2)
    & = S_{\mathrm{top},x}(j_1,j_2) + S_{\mathrm{top},y}(j_1,j_2).
  \end{align}
\end{subequations}

\subsection{Time Evolution and Majorana Correlation Matrix}

In order to calculate the string order parameter, we first introduce a
$2N\times2N$ skew-symmetric Majorana correlation matrix $C(t)$ whose elements
are defined by
\begin{equation}
  \label{eq:C}
  C_{kk'}(t)
  = -i\left(\Braket{\gamma_k(t) \gamma_{k'}(t)} - \delta_{kk'}\right)
\end{equation}
for $k,k'=1,\cdots,2N$.
% Note that this matrix is anti-symmetric: $C_{kk'}=-C_{k'k}$.
%
The Hamiltonian (\ref{eq:model}) is quadratic and one can apply Wick's theorem to evaluate the string order
parameter. Wick's theorem tells us that the expectation value of the
products of $\gamma_k$ is the sum of all the possible products of the
expectation values of pairs. The permutation between operators can change the
overall sign if it happens in an odd number. The results can thus be expressed
in terms of Pfaffian of the Majorana correlation matrix
\cite{Barouch1971feb,Calabrese2012julb}. More explicitly,
\begin{equation}
  S_{\mathrm{top},x}(j_1,j_2) = \operatorname{Pf}[C^{(x)}(j_1,j_2)],
  \quad
  S_{\mathrm{top},y}(j_1,j_2) = - \operatorname{Pf}[C^{(y)}(j_1,j_2)],
\end{equation}
where $C^{(x)}(j_1,j_2)$ and $C^{(y)}(j_1,j_2)$ are
$2(j_2-j_1)\times2(j_2-j_1)$ submatrices of $C$ formed by rows and columns
$2j_1,2j_1+1,\cdots,2j_2-2,2j_2-1$ and
$2j_1-1,2j_1+1,2j_1+2,\cdots,2j_2-3,2j_2-2,2j_2$, respectively.

Now we formulate the time evolution of the correlation matrix $C(t)$. The
Heisenberg equations of motion for the Majorana operators lead to the differential equation for $\tC_{kk'}(t) = \Braket{\gamma_k(t) \gamma_{k'}(t)}$:
\begin{equation}
  i\hbar\frac{d\tC(t)}{dt}
  = M(t) \tC(t) - \tC(t) M(t)
\end{equation}
with
\begin{equation}
  M(t)
  =
  i
  \left[
    \begin{array}{cccccccccc}
      0 & -\mu_1 & 0 & w_- & 0 & 0 & \cdots & 0 & 0 & 0
      \\
      \mu_1 & 0 & w_+ & 0 & 0 & 0 & \cdots & 0 & 0 & 0
      \\
      0 & -w_+ & 0 & -\mu_2 & 0 & w_- & \cdots & 0 & 0 & 0
      \\
      -w_- & 0 & \mu_2 & 0 & w_+ & 0 & \ddots & 0 & 0 & 0
      \\
      0 & 0 & 0 & -w_+ & 0 & -\mu_3 & \ddots & w_- & 0 & 0
      \\
      0 & 0 & -w_- & 0 & \mu_3 & 0 & \ddots & 0 & 0 & 0
      \\
      \vdots & \vdots & \vdots & \ddots & \ddots &
      \ddots & \ddots & -\mu_{N{-}1} & 0 & w_-
      \\
      0 & 0 & 0 & 0 & -w_- & 0 & \mu_{N{-}1} & 0 & w_+ & 0
      \\
      0 & 0 & 0 & 0 & 0 & 0 & 0 & -w_+ & 0 & -\mu_N
      \\
      0 & 0 & 0 & 0 & 0 & 0 & -w_- & 0 & \mu_N & 0
    \end{array}
  \right]
\end{equation}
and
\begin{equation}
  w_\pm \equiv \frac{\Delta \pm w}{2}.
\end{equation}
By solving the differential equation, we obtain
\begin{equation}
  \tC(t) = U(t) \tC(0) U^t(t)
\end{equation}
with the time evolution matrix
\begin{equation}
  U(t)
  = T \exp\left[-\frac{i}{\hbar} \int_0^t d\tau\, M(\tau)\right].
\end{equation}
Finally,
\begin{equation}
  C(t)
  = (-i)(\tC(t) - 1)
  = (-i)(U(t) \tC(0) U^t(t) - 1)
  = U(t) C(0) U^t(t)
\end{equation}
since $U(t)$ is orthogonal matrix. Note that if $C(0)$ is real, the $C(t)$ is
real for all time $t$.

\subsection{General Structure of Majorana Correlation Matrix}

In general, the wave functions of the two Majorana edge states overlap with each other, giving rise to a finite energy
splitting (although exponentially small for a long chain). So the ground state
has a definite fermion parity: In our study we choose $P
= 1$. Note that the fermion parity, \eqnref{eq:parity}, is no other than the
Pfaffian of $C$: $P = \operatorname{Pf}[C]$. Therefore, $\operatorname{Pf}[C] = 1$ throughout the
time evolution.
On the other hand, the real $2N\times2N$ skew-symmetric matrix can be always
block-diagonalized by an orthogonal matrix $V$: $C = V D V^t$ with
\begin{equation}
  D
  =
  \bigoplus_{n=1}^N
  \begin{bmatrix}
    0 & \lambda_n
    \\
    -\lambda_n & 0
  \end{bmatrix}
\end{equation}
where $\lambda_n$ are real. Since $\operatorname{Pf}[C] = \det[V]\,
\operatorname{Pf}[D] = \operatorname{Pf}[D] = \prod_{n=1}^N \lambda_n = 1$ and
the correlation between the Majorana operators cannot be larger than 1 in
magnitude, one can conclude that $\lambda_n = 1$ for all $n$.  It leads to an
interesting physical implication: In a proper basis, each of $2N$ Majorana
fermions forms a pair with another single Majorana fermion (not a superposition
of pairs with different Majorana fermions), resulting in $N$ definite pairs.
The simplest examples are shown in \figsref{fig:Majoranachain}(b) and (c); with
the fermion parity fixed, two edge Majorana fermions, isolated otherwise,
should form a non-local pair.  It is stressed that for a general dynamical
wavefunction the Majorana fermions are not necessarily localized at one site as
in \figsref{fig:Majoranachain}(b) and (c), but they can move along the wire or
disperse with time. Such propagation and spread of each Majorana fermion are
reflected in the time evolution matrix $U(t)$.  In other words, by tracking
down the time evolution of each column of $U(t)$, one can find that which
Majorana fermion forms a pair with which Majorana fermion at each
time. Importantly, this information in turn can be used to interpret the
dynamical change of $S_\mathrm{top}$ with time. For example,
$S_{\mathrm{top},x}(j_1,j_2)$ would be maximal when the Majorana fermions
$\gamma_{2j_1}, \gamma_{2j_1+1}, \cdots, \gamma_{2j_2-2}, \gamma_{2j_2-1}$
strongly form pairs by themselves. If any of them is bound to that outside the
region, then the topological order in that region must decrease.

\subsection{Quench Dynamics}

Throughout the study, for simplicity, we take the Ising limit in which the
$p$-wave superconducting order parameter $\Delta$ is equal to the hopping
amplitude $w$: $w_+ = w$ and $w_- = 0$. The Ising case can reveal the key
essence of the dynamics of the topological order.  Since we are interested in
the quench dynamics of the system, the parameters of the system are driven to
change in time. Taking into account the feasibility of experimental
realization, the hopping amplitude as well as the superconducting gap is fixed
in time and position-independent, while the position-dependent chemical
potential $\mu_j(t)$, which can be easily controlled experimentally, is varied with time so
that the whole or a part of the system experience dynamical topological phase
transition.

In our study we consider two cases: In the first case, the wire is uniform so
that $\mu_j(t) = \mu(t)$ and a quantum quench is applied at time $t=0$
\begin{equation}
  \label{eq:casei}
  \mu(t)
  =
  \begin{cases}
    0, & t < 0
    \\
    \mu_f, & t > 0
  \end{cases}
\end{equation}
so that the whole wire is driven from the deep topological phase ($\mu = 0$) to
the non-topological phase ($\mu = \mu_f > w$).
In the second case, only the central region with $N_\mathrm{sys}$ sites (called
as the system) is initially prepared in the topological phase ($\mu_j =
\mu_\mathrm{sys} = 0$) and the side regions (called as the environment) with
$N_\mathrm{env} \equiv (N-N_\mathrm{sys})/2$ sites in each side are in the
non-topological phase ($\mu_j = \mu_\mathrm{env} > w$). At time $t=0$, the
system part is driven to the non-topological phase like in the environment:
\begin{equation}
  \label{eq:caseii}
  \mu_\mathrm{sys}(t)
  =
  \begin{cases}
    0, & t < 0
    \\
    \mu_f, & t > 0
  \end{cases}
  \quad\text{and}\quad
  \mu_\mathrm{env}(t) = \mu_f.
\end{equation}
For convenience, we set $\hbar = 1$ and focus on the zero temperature case.

\section{Time Evolution of Topological Order in Uniform Wire}
\label{sec:uniform}

\begin{figure}[!b]
  \centering
  \includegraphics[width=.9\textwidth]{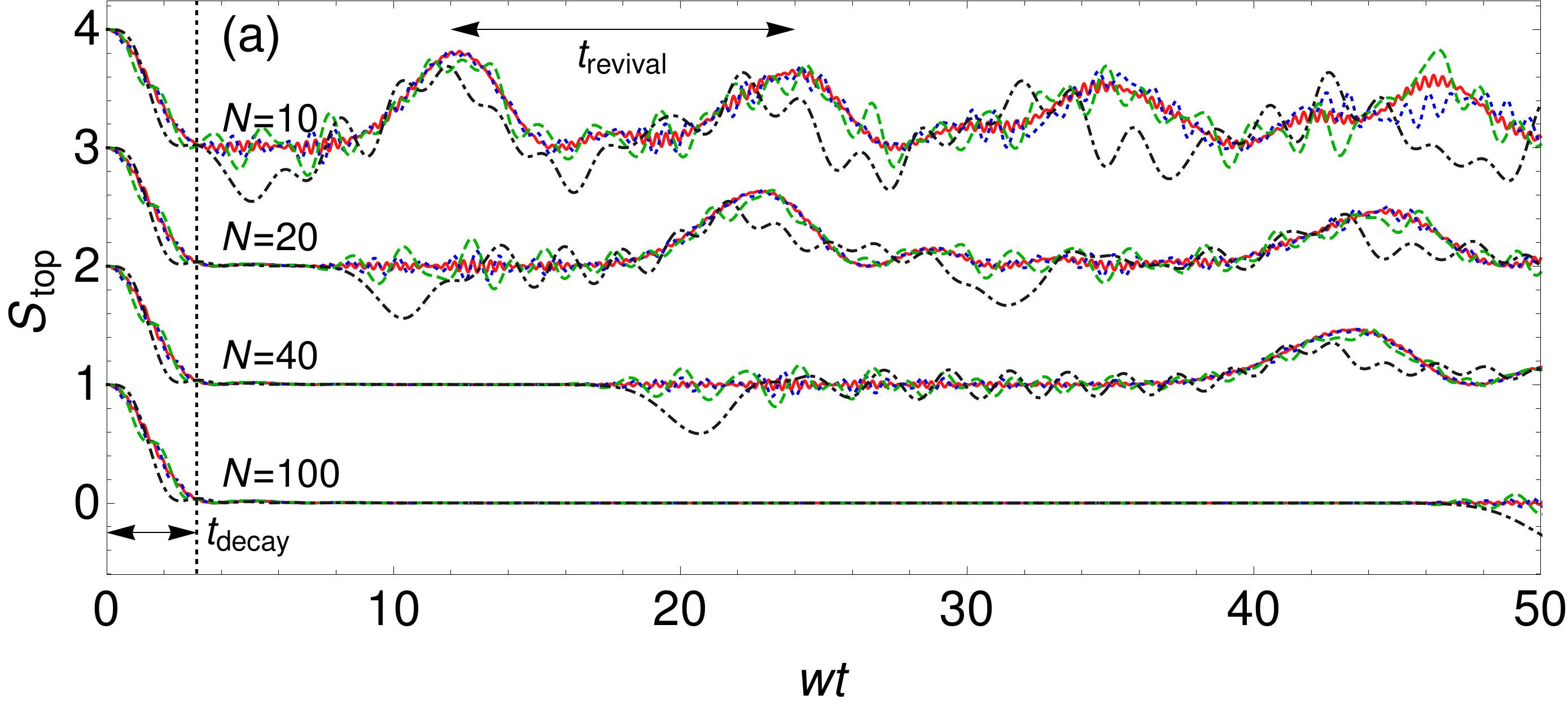}\\
  \includegraphics[width=.45\textwidth]{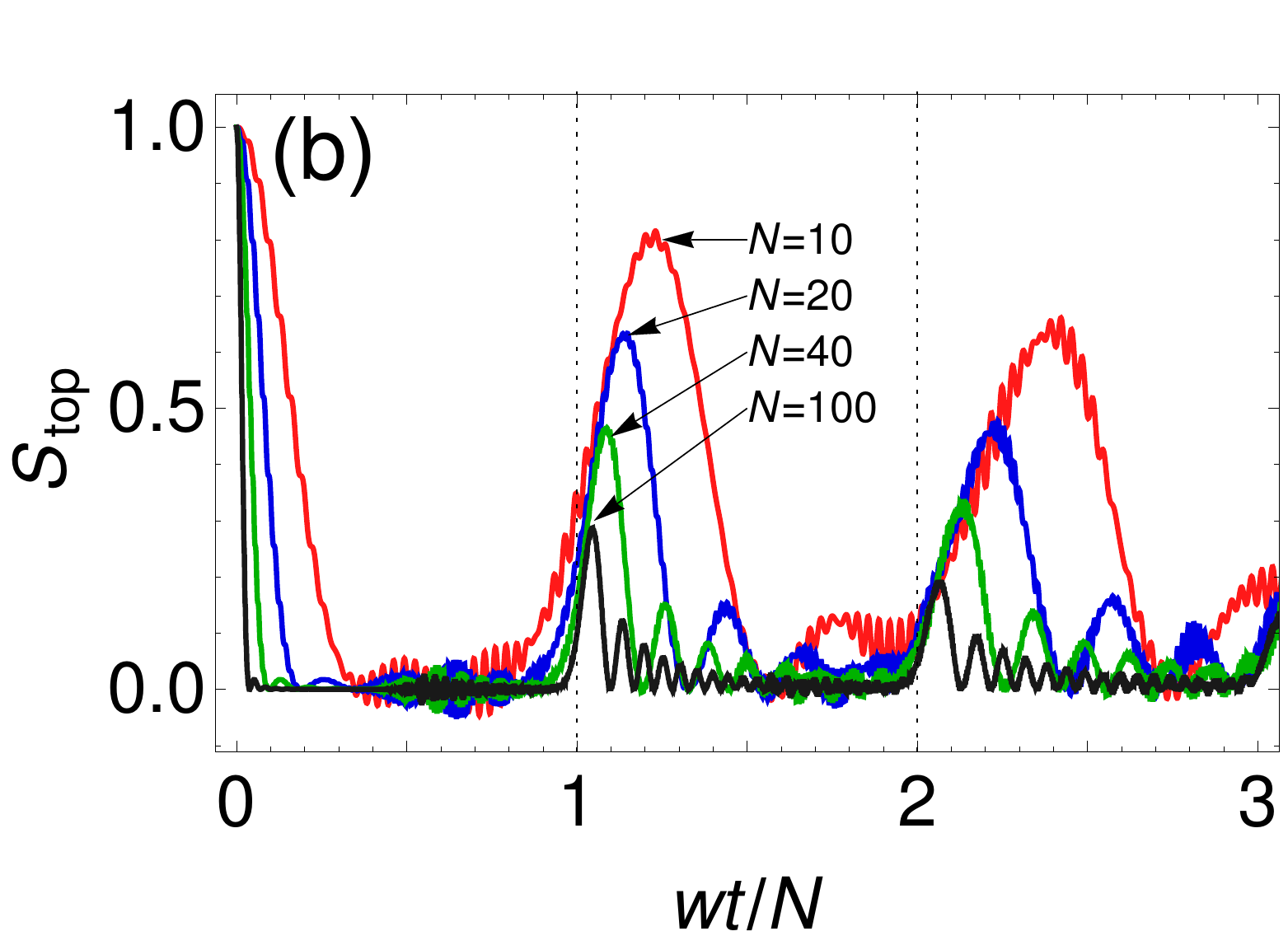}%
  \includegraphics[width=.45\textwidth]{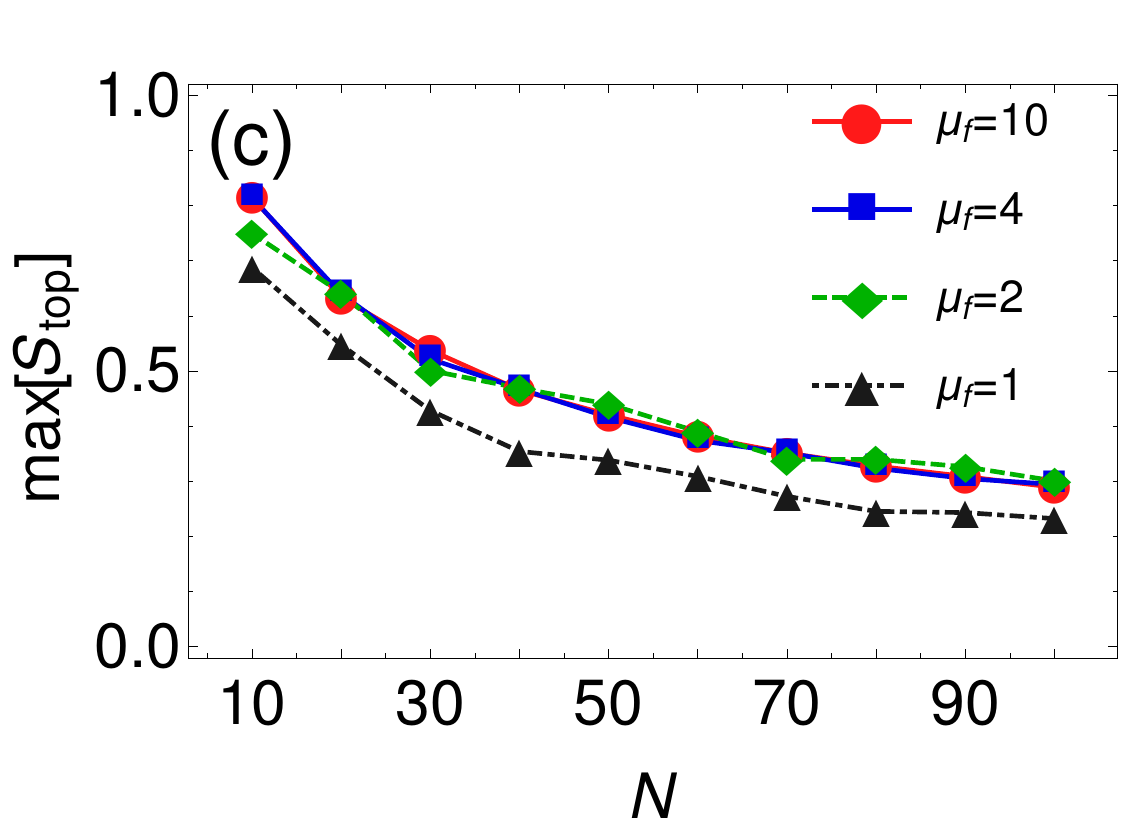}%
  \caption{(a) Time evolution of string order parameter $S_\mathrm{top}$ for
    topological superconducting wires with different lengths after the quench,
    \eqnref{eq:casei}, for several values for $\mu_f/w$: 10 (red solid), 4
    (green dotted), 2 (blue dashed), and 1 (black dot-dashed), (b) Time-scaled
    version of the time evolution for $\mu_f/w = 10$, and (c) the maximum
    amplitude of $S_\mathrm{top}$ at the first revival as a function of the wire
    length $N$. Note that the curves for different system size $N$ in figure
    (a) are shifted by one. Here we set $w = \Delta = 1$.}
  \label{fig:sop:casei}
\end{figure}

In this section we consider the quantum quench of a uniform $p$-wave
superconducting wire from the topological to non-topological phase according to
\eqnref{eq:casei}.  \Figref{fig:sop:casei} displays the time evolution of
$S_\mathrm{top}$ after the quantum quench at $t=0$. Here we summarize the two
characteristic behaviors of the string order parameter uncovered in
Fig.~\ref{fig:sop:casei} (and below we explain them in terms of the motion of
Majorana fermions):
(i) The topological orders defined by \eqnsref{eq:sop:x}, (\ref{eq:sop:y}), and
(\ref{eq:sop}) do not die away immediately after the quench, but instead decay
with time before they vanish completely. Interestingly, the shape of
$S_\mathrm{top}(t)$ during the decay and the decaying time $t_\mathrm{decay}$
are quite universal: As long as $\mu_f$ is not too close to the transition
point $w$, we numerically find $w t_\mathrm{decay} \approx \pi$, which is
immune to the wire length $N$ and the final value of the chemical potential
$\mu_f$. This size-independence is quite interesting considering that the
topological order is based on the nonlocal products of operators [see
\eqnsref{eq:sop:x} and (\ref{eq:sop:y})]. We also find that as long as $\mu_f >
w$ the overall behavior of $S_\mathrm{top}$ is qualitatively same for all
values of $\mu_f$, except additional small oscillations whose amplitude and
period decrease with increasing $\mu_f$, as seen in \figref{fig:sop:casei}(a).
(ii) The topological order is revived repeatedly although the amplitude
gradually decreases. Specifically, the topological order reappears at $t
\approx n N/w$ ($n=1,2,\cdots$) and is kept for a duration time which is around
$2t_\mathrm{decay}$, forming peaks centered at $t \approx nN/w +
t_\mathrm{decay}$, as seen in \figsref{fig:sop:casei}(a) and (b).  The revival
period, the time between the peaks in $S_\mathrm{top}(t)$, is definitely
related to the system size: $t_\mathrm{revival} \sim N/w$. Surely, the revival
is the finite-size effect. \Figref{fig:sop:casei}(c) shows that the maximum
amplitude of $S_\mathrm{top}$ at its first revival decreases with the system
size and that its dependence on $N$ is immune to the quench strength as long as
$\mu_f$ is sufficiently larger than $w$. This observation can be explained in
terms of the dispersion of the Majorana wave function, which will be discussed
in greater detail later.
Note that a similar kind of revival of Majorana physics has been reported by
examining the survival probability of edge Majorana states in the same system
\cite{Rajak2014apr}. In their study the survival of local Majorana wave function
is maximal in the quench to the transition point ($\mu_f = w$). However, the
revival of non-local topological order in our study turns out to be quite
robust no mater what value $\mu_f$ has.

\begin{figure}[!t]
  \centering
  \includegraphics[width=.66\textwidth]{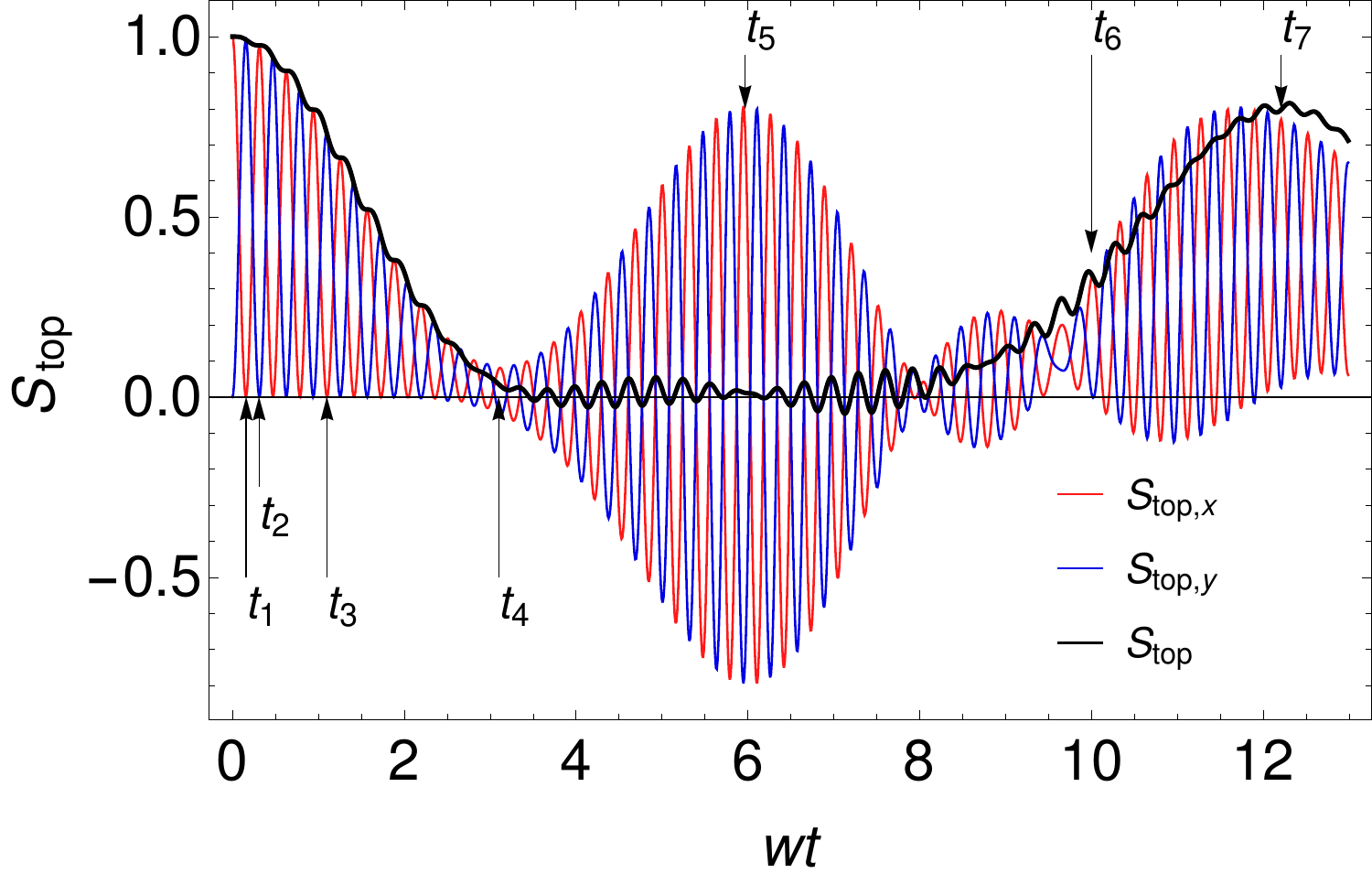}
  \caption{Time evolution of string order parameters $S_{\mathrm{top},x/y}$ and
    $S_\mathrm{top}$ for $N=10$ topological superconducting wires after the
    quench, \eqnref{eq:casei} with $\mu_f/w = 10$.}
  \label{fig:sop:xy}
\end{figure}

In passing, we examine and compare the time evolutions of different string
order parameters, $S_{\mathrm{top},x/y}$ as well as $S_\mathrm{top}$, more
closely.  \Figref{fig:sop:xy} shows that unlike $S_\mathrm{top}$ the string
order parameters $S_{\mathrm{top},x/y}$ oscillate rapidly in the
anti-synchronized way. The initial oscillation with $\pi/2$ phase difference
between $S_{\mathrm{top},x/y}$ can be understood in the spin
language. Initially, the ferromagnetic coupling $J_x = w/2$ aligns all the
spins in the $x$ direction. Turning on the transverse magnetic field $h_z$
makes the spins precess in the $xy$ plane, transferring the correlation in the
$x$ direction into that in the $y$ direction and vice versa. The spin
precession under the transverse field is responsible for the oscillation whose
period is then inversely proportional to $1/|h_z| \propto 1/\mu_f$.

\begin{figure}[!t]
  \centering
  \includegraphics[width=\textwidth]{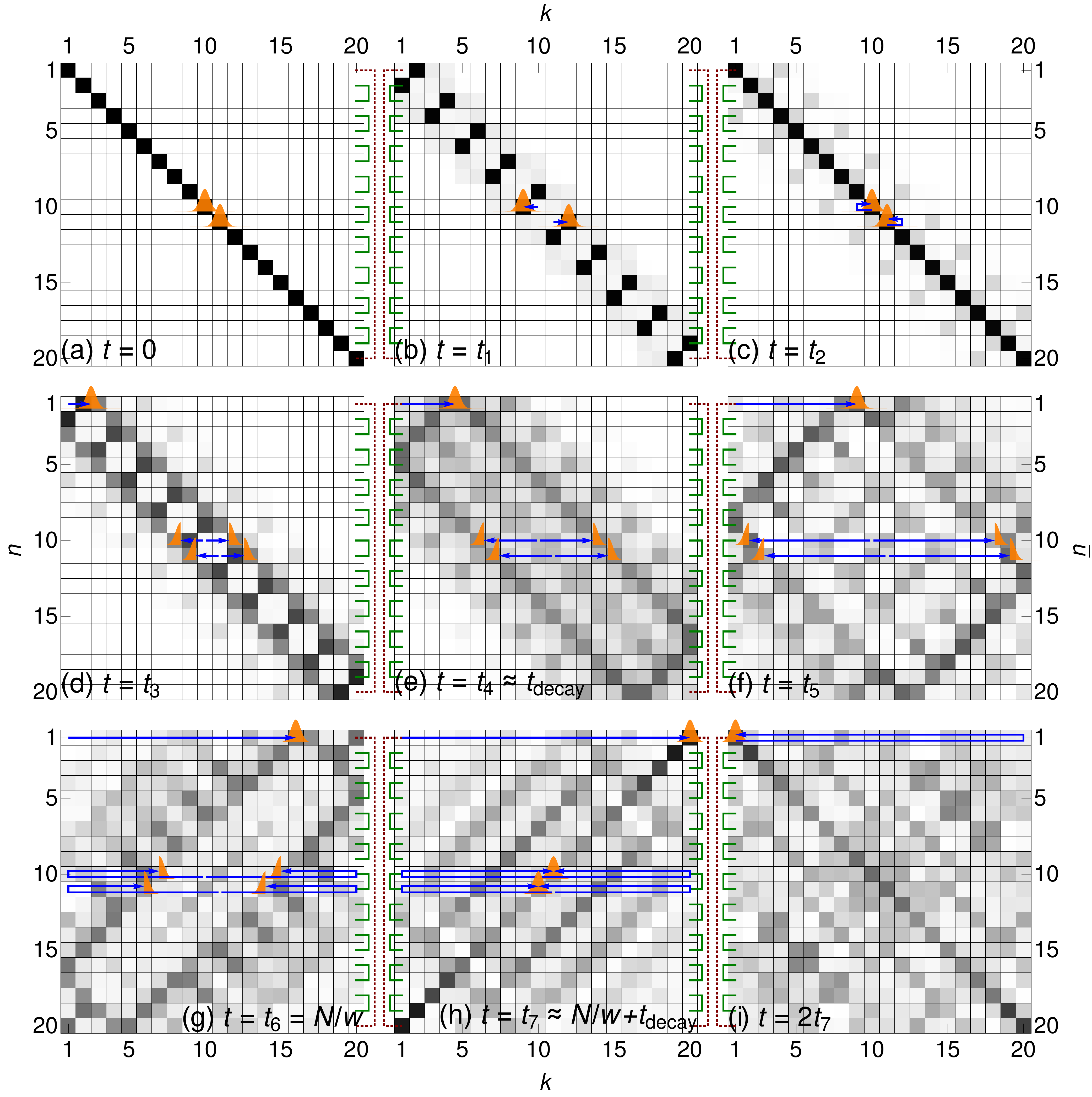}
  \caption{Density plot of the time evolution matrix $|U_{kn}|$ for the time
    evolution of $N=10$ topological superconducting wires after the quench,
    \eqnref{eq:casei} with $\mu_f/w = 10$. The times at which the matrix
    elements are taken are indicated in \figref{fig:sop:casei}: (a) $t=0$, (b)
    $t=t_1$ when $S_{\mathrm{top},x}=0$ for the first time, (c) $t=t_2$ when
    $S_{\mathrm{top},y}=0$ for the second time, (d) $t=t_3$ when
    $S_{\mathrm{top},x}=0$, (e) $t=t_4\approx t_\mathrm{decay}$ when
    $S_\mathrm{top}$ vanishes almost, (f) $t=t_5$ when the oscillation of
    $S_{\mathrm{top},x/y}$ is maximal again, (g) $t=t_6=N/w$, (h) $t=t_7\approx
    N/w+t_\mathrm{decay}$ when $S_\mathrm{top}$ is fully restored, and (i)
    $t=2t_7$. The green and red lines connecting rows indicates the pairing
    between the pairs of Majorana states.}
  \label{fig:sop:casei:cm}
\end{figure}

All those aspects of the change of the string order parameter with time can be
explained in terms of the time evolution of Majorana wave function reflected in
the matrix $U(t)$, as we discuss in the remaining of the section.
\Figref{fig:sop:casei:cm} displays the time evolution of Majorana wave
function:
Each row of the density plots corresponds to the wave function of the
Majorana state ordered in such a way that at $t=0$ the $n^\mathrm{th}$ state
($1\le n\le 2N$) is localized at Majorana site $k = n$. Hence, according to
the initial pairing between Majorana fermions depicted in
\figref{fig:Majoranachain}(b), the $n=2j$ and $2j+1$ Majorana states
($j=1,2,\cdots,N-1$) are paired and the edge Majorana states ($n=1$ and $2N$)
are also paired due to the fermion parity fixing [see the green and red lines
in \figref{fig:sop:casei:cm}(a)]. While the wave functions of the Majorana
states change with time, theses pairings are maintained throughout the time
evolution.
Turning on the chemical potential couples $\gamma_{2j-1}$ and $\gamma_{2j}$ so
that the Majorana fermions hop between two Majorana sites in a real site
$j$. At $t=t_1\approx\pi/2\mu_f$, the Majorana fermions in a real site exchange
between them [see \figref{fig:sop:casei:cm}(b)] so that the Majorana
configuration is almost same as \figref{fig:Majoranachain}(c), resulting in
$S_{\mathrm{top},x} = 0$ and $S_{\mathrm{top},y} \approx 1$.
At $t=t_2\approx\pi/\mu_f$, the Majorana fermions return to their starting
points [see \figref{fig:sop:casei:cm}(c)], restoring $S_{\mathrm{top},x} \approx
1$ almost with $S_{\mathrm{top},y} = 0$.

The oscillation is accompanied with the dispersion of the Majorana states,
making the pairing between the bulk and edge Majorana states gradually
increasing and accordingly weakening $S_\mathrm{top}$. On top of it, each Majorana wave function
(except $n=1,2N$ Majorana states) is split into two propagation modes, moving
in opposite directions [see
\figsref{fig:sop:casei:cm}(d) and (e)].
At $t=t_4\approx\pi/w$ [see \figref{fig:sop:casei:cm}(e)], the initial edge
Majorana states ($n=1,2N$) enter into the bulk completely, and the propagation
of split Majorana states, initially starting in the bulk, hits the ends of the
chain so that the correlation between the bulk and edge Majorana fermions\
becomes maximal, resulting in the vanishing of $S_{\mathrm{top},x/y}$ and $S_\mathrm{top}$. Since the decaying of the topological order happens when the initial
edge Majorana states enter into the bulk, the decaying time does not depend on
the system size. Also, noting that the propagation speed of the Majorana states
is mainly determined by the hopping amplitude $w$, the chemical potential
strength does not affect the decaying time $t_\mathrm{decay}$.

For $t \lesssim N/w$, the topological order parameters $S_\mathrm{top}$ and
$S_{\mathrm{top},x/y}$ remain being suppressed except around $t \sim N/2w$ in
which $S_{\mathrm{top},x}$ and $S_{\mathrm{top},y}$ oscillate out of phase. A typical
distribution of Majorana states in that period is shown in
\figref{fig:sop:casei:cm}(f). One can see that the Majorana states for $n =
8\sim13$ become mostly localized at the boundaries and form pairs by
themselves: $\gamma_1$ and $\gamma_{2N}$, and $\gamma_2$ and $\gamma_{2N-1}$,
simultaneously.  The other Majorana states inside the bulk are binding by
themselves as well. It satisfies the condition to have finite $S_{\mathrm{top},x/y}$ as shown in \figref{fig:sop:xy}. Note that the pairing inside
the bulk is not like those in \figref{fig:Majoranachain}(b) nor (c), where the
Majorana fermions in neighboring sites form pairs, but instead is more like
that in \figref{fig:Majoranachain}(a), where the Majorana fermions in the same
site is bound. Hence, it results in the almost perfect cancellation between
$S_{\mathrm{top},x}$ and $S_{\mathrm{top},y}$, making $S_\mathrm{top}$ vanish. In this
respect, $S_\mathrm{top}$, not $S_{\mathrm{top},x}$ nor $S_{\mathrm{top},y}$, is the
more adequate order parameter to measure the topological order.

At $t = t_6 = N/w$ [see \figref{fig:sop:casei:cm}(g)], the front part of the
initial edge Majorana wave function $(n = 1, 2N)$ starts to reach the opposite
ends, and the bulk Majorana states which was split becomes reunited. That is,
apart from the dispersion of Majorana states, the Majorana configuration starts
to be restored to the initial one with reverse ordering, which accordingly
revives the topological order. At $t = t_7 \approx N/w + t_\mathrm{decay}$ [see
\figref{fig:sop:casei:cm}(h)], the topological order is maximally restored,
restoring the initial Majorana configuration.  Note that while the bulk states
are considerably dispersed, the Majorana states for $n=1,2N$, forming a strong
pair, are still quite localized at the edges. This is the reason why the
restoration of $S_\mathrm{top}$ is quite strong.
Finally, at $t = 2t_7 \approx 2N/w$ [see \figref{fig:sop:casei:cm}(i)], the
Majorana configuration returns to the initial one with much dispersion, leading
to the second revival of the topological order. Note that our analysis
  which tracks down the pairing between Majorana states explains why the
  revival period of the topological order is $N/w$, the half of the round-trip
  period ($=2N/w$) of Majorana states.

\begin{figure}[!t]
  \centering
  \includegraphics[width=.7\textwidth]{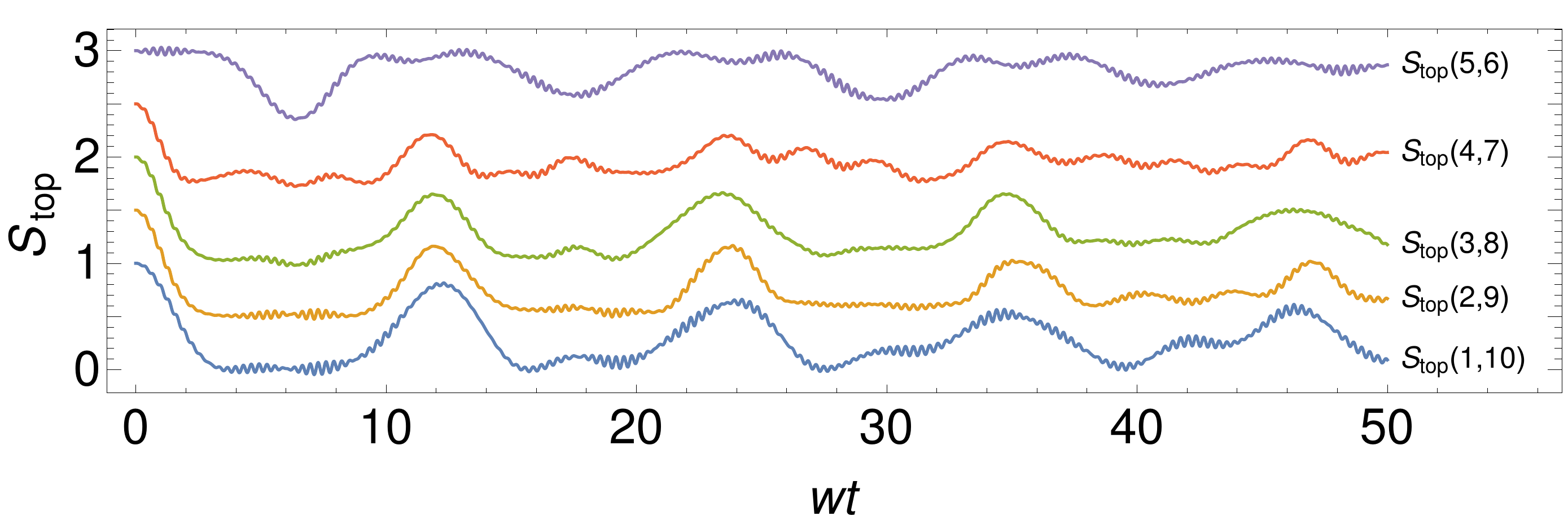}
  \caption{Time evolution of string order parameter $S_\mathrm{top}(j_1,j_2)$ for
    $N=10$ topological superconducting wires after the quench,
    \eqnref{eq:casei} with $\mu_f/w = 10$. Each curve is shifted by 1/2.}
  \label{fig:sop:casei:part}
\end{figure}

Now we examine the time evolution of $S_\mathrm{top}(j_1,j_2)$ for central
segments of the wire from site $j_1$ to site
$j_2=N+1-j_1$. \Figref{fig:sop:casei:part} shows that the revival of the
topological order happens for all $S_\mathrm{top}(j_1,j_2)$ simultaneously. It
means that the pairing of Majorana fermions is like that in
\figsref{fig:Majoranachain}(b) or (c) in any length scale, once the order is
restored. For the quench in uniform wires, this is the case as can be seen in
\figsref{fig:sop:casei:cm}(h) and (i).

Finally, we address interesting relations between the string order parameters
$S_{\mathrm{top},x/y}$ and the elements of the correlation matrix $C$. From the
observations up to now, it is obvious that $S_{\mathrm{top},x}$ increases as
the Majorana fermions $\gamma_k$ on sites $k=2,\cdots,2N-1$ form strong pairs
by themselves. Under the condition that \emph{the fermion parity is fixed},
every Majorana fermion should be bound to one of others. Hence, with
$S_{\mathrm{top},x}$ close to one by binding $\gamma_k$ for $k=2,\cdots,2N-1$,
a strong pair between $\gamma_1$ and $\gamma_{2N}$ should be formed. It implies
that $S_{\mathrm{top},x}$ and $C_{1,2N}$ are positively correlated. A similar
argument also leads to a positive correlation between $S_{\mathrm{top},y}$ and
$C_{2,2N-1}$. As proven in \ref{sec:app:topologicalorder}, this kind of
correlations is exact:
\begin{subequations}
  \begin{align}
    \label{eq:C:stopx}
    S_{\mathrm{top},x} & = C_{1,2N}
    \\
    \label{eq:C:stopy}
    S_{\mathrm{top},y} & = -C_{2,2N-1}
  \end{align}
\end{subequations}
[see \eqnsref{eq:C:stopx:proof} and (\ref{eq:C:stopy:proof})]. Therefore, the
topological order can be obtained simply from the correlations between edge Majorana fermions.
This result may lead to a false understanding that the topological order is in
fact local since $C_{1,2N}$ or $C_{2,2N-1}$ are local correlations. However, we
can argue that the topological order is really nonlocal in two ways:
(i) These relations, \eqnsref{eq:C:stopx} and (\ref{eq:C:stopy}) are based on
the assumption that the fermion parity is fixed. However, as pointed out in
Ref.~\cite{Bahri2014apr}, the string order parameter is insensitive to the
linear combination of degenerate ground states with different parities. So even when the relations between the string order parameter and the edge
Majorana correlation are not valid, the nonlocal string order parameter is
still unambiguously defined.
(ii) The derivation leading to \eqnsref{eq:C:stopx} and (\ref{eq:C:stopy})
states that the string order parameter over a part of the system is related to
the correlation outside the part [see
\eqnref{eq:stop:general}]. \Eqnsref{eq:C:stopx} and (\ref{eq:C:stopy}) are
special cases where the remaining part consists of only two (edge) Majorana
fermions. So the topological order is truly global property of the system,
reflecting the correlations all over the system.

\section{Propagation of Topological Order into Non-Topological Regions}
\label{sec:propagation}

\begin{figure}[!t]
  \centering
  \includegraphics[width=.5\textwidth]{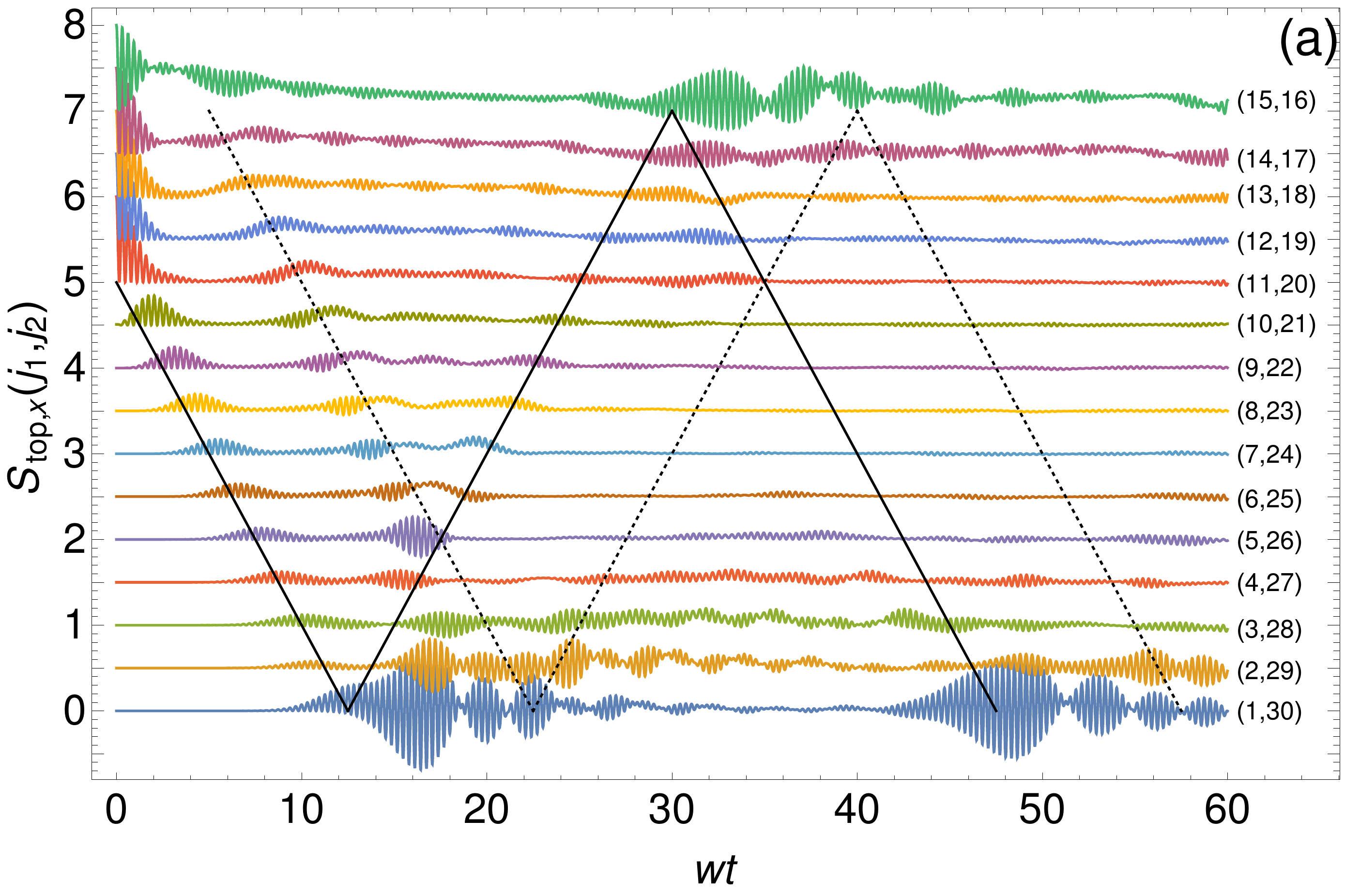}%
  \includegraphics[width=.5\textwidth]{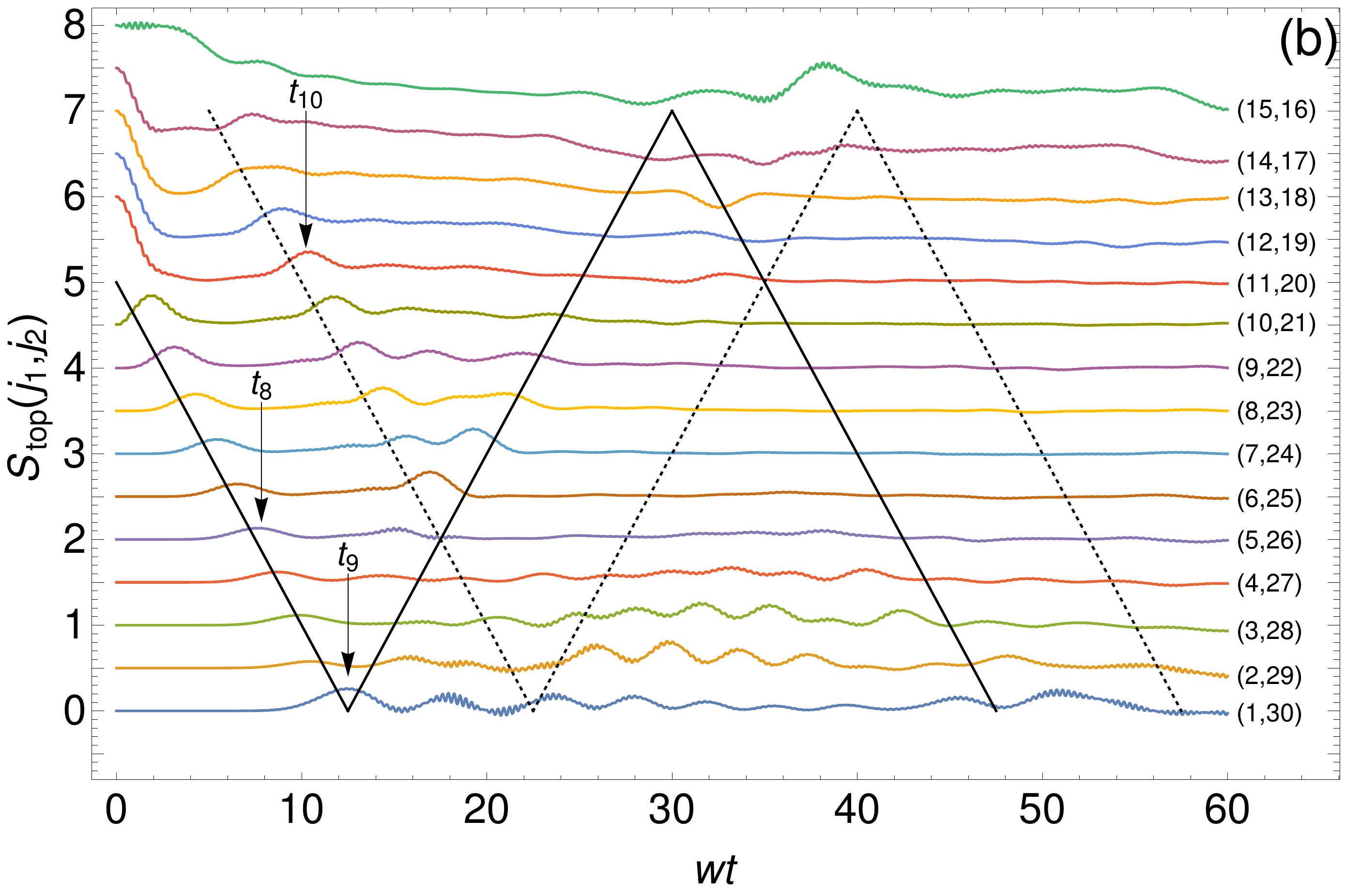}
  \caption{Time evolution of string order parameter (a)
    $S_{\mathrm{top},x}(j_1,j_2)$ and (b) $S_\mathrm{top}(j_1,j_2)$ for $N=30$
    topological superconducting wires with $N_\mathrm{sys}=10$ central sites
    initially prepared in topological phase after the quench,
    \eqnref{eq:caseii} with $\mu_f/w = 10$. Each curve is tagged with the
    corresponding site interval $(j_1,j_2{=}N{+}1{-}j_1)$ and is shifted by 1/2
    for clarity.  The solid and dotted lines are guide for Majorana fermion
    propagation which is explained in the text.}
  \label{fig:sop:caseii}
\end{figure}

Now we consider the case in which only a central part (``system'' with
$N_\mathrm{sys}$ sites) of the superconducting wire is initially in topological
phase and the outer part (``environment'' with $N_\mathrm{env}$ sites on the
left and another $N_\mathrm{env}$ sites on the right) is in non-topological
phase, according to \eqnref{eq:caseii}. \Figref{fig:sop:caseii} shows the time
evolution of $S_{\mathrm{top},x}(j_1,j_2)$ and $S_\mathrm{top}(j_1,j_2)$ for
every (centered) region (we take $j_1$ and $j_2$ symmetrically so that
$j_2=N+1-j_1$) of the wire with $N=30$ and $N_\mathrm{sys}=10$ (accordingly,
$N_\mathrm{env}=10$) after a sudden quench.
At $t=0$, the system part is in topological phase so that $S_{\mathrm{top},x}(j_1,j_2) = S_\mathrm{top}(j_1,j_2) = 1$ for $N_\mathrm{env}<j_1\le
N/2$. However, the string order parameters for larger regions which extend into
the environment part are initially zero: $S_{\mathrm{top},x}(j_1,j_2) = 0$ for $j_1 \le
N_\mathrm{env}$ since $\gamma_{2j_1}$ at site $j_1$ in the environment region form
a strong pair with $\gamma_{2j_1-1}$ which is not included in the string of
Majorana operators $\gamma_{2j_1} \gamma_{2j_1+1} \cdots \gamma_{2j_2-1}$ for
$S_{\mathrm{top},x}(j_1,j_2)$ [see \eqnref{eq:sop:x:region}].

\begin{figure}[!t]
  \centering
  \includegraphics[width=.5\textwidth]{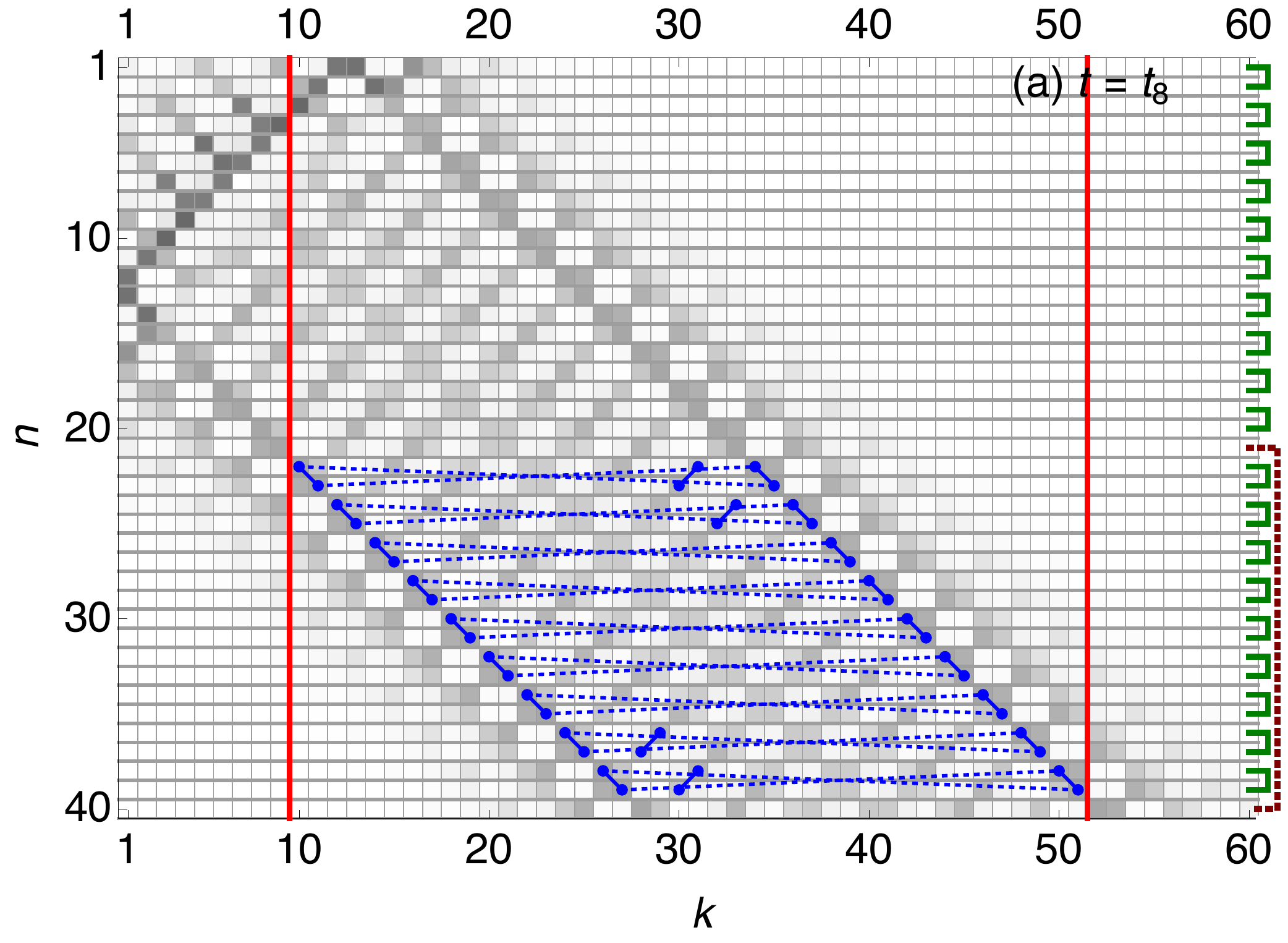}%
  \includegraphics[width=.5\textwidth]{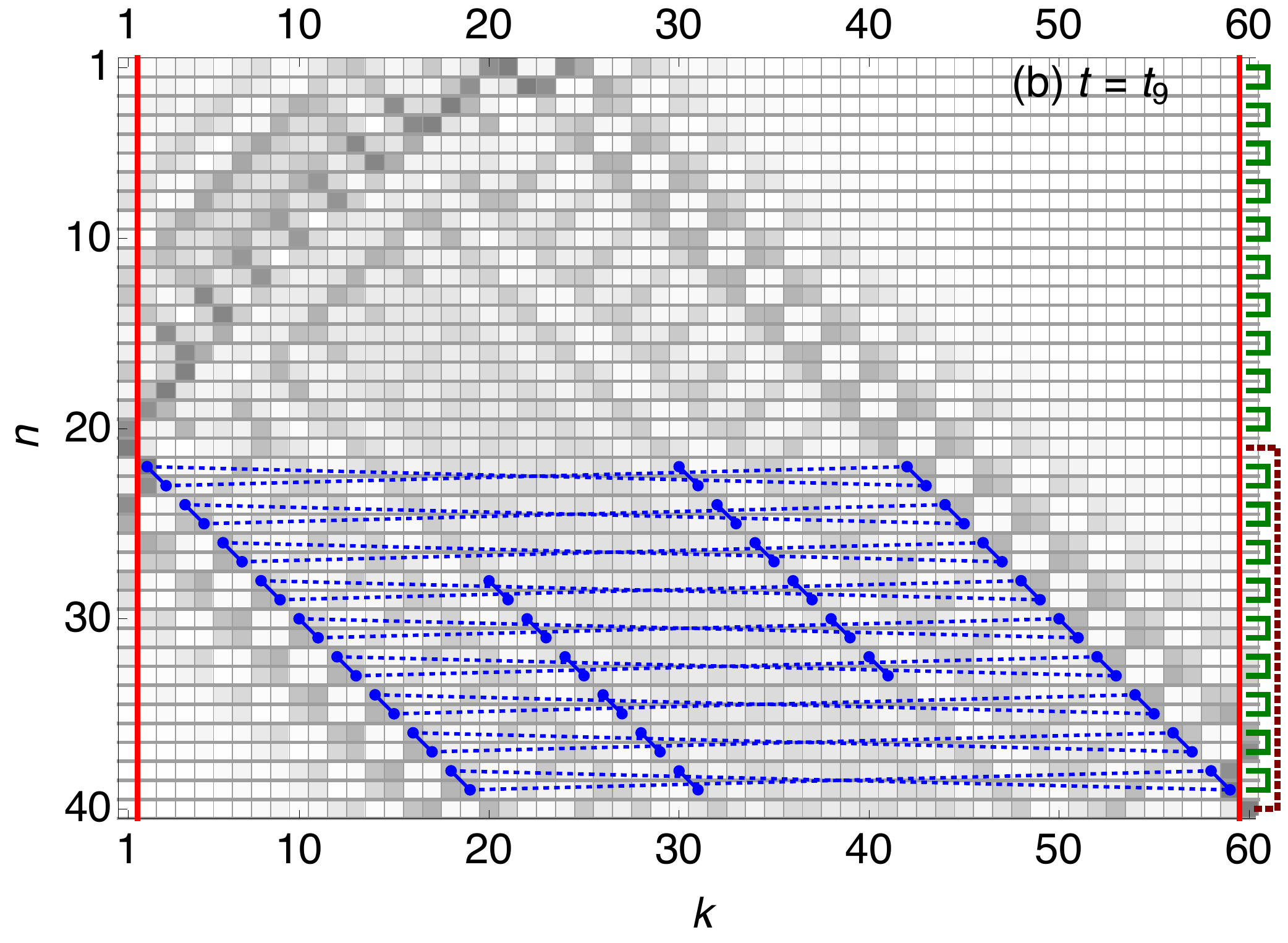}\\
  \includegraphics[width=.5\textwidth]{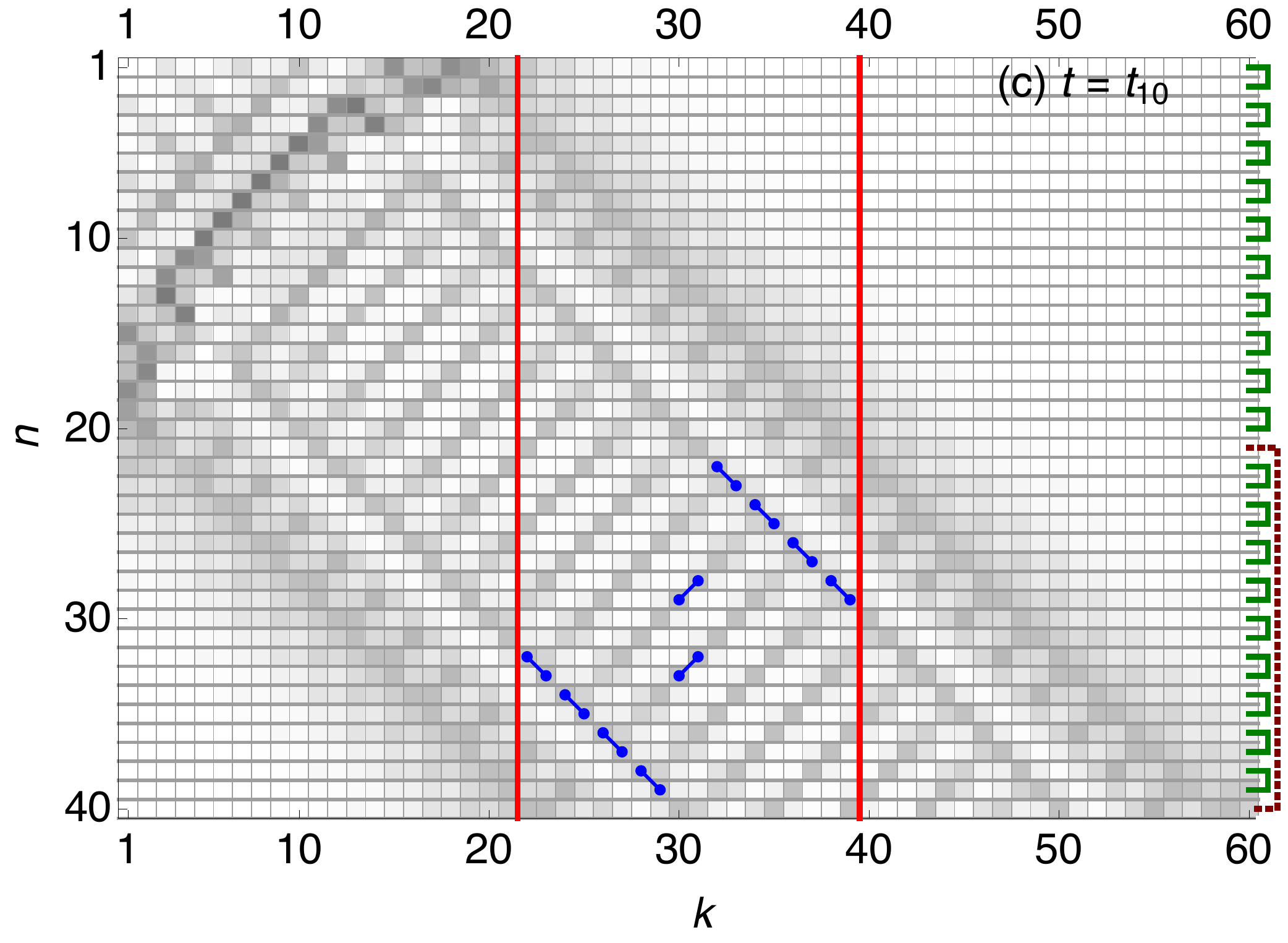}\\
  \caption{Density plot of the time evolution matrix $|U_{kn}|$ for the time
    evolution of $N=30$ and $N_\mathrm{sys}=10$ topological superconducting
    wires initially prepared as described in \figref{fig:sop:caseii}. The times
    at which the matrix elements are taken are indicated in
    \figref{fig:sop:caseii}: (a) $t=t_8$ when $S_{\mathrm{top}}(5,26)$ becomes
    maximal for the first time, (b) $t=t_9$ when $S_{\mathrm{top}}(1,30)$
    becomes maximal for the first time, and (c) $t=t_{10}$ when
    $S_{\mathrm{top}}(11,20)$ is fully revived for the first time. In all
    cases, $S_{\mathrm{top},y} \approx 0$. The green and red lines connecting
    rows (in the right side of the figures) indicates the pairing between the
    pairs of Majorana states. The red vertical lines indicate the regions in
    which the topological order is measured. The blue small circles and (solid
    and dotted) lines connecting them pick the Majorana pairs which mostly
    contributes to the corresponding $S_{\mathrm{top},x}$.}
  \label{fig:sop:caseii:cm}
\end{figure}

After quench, $S_{\mathrm{top},x}(j_1,j_2)$ and $S_\mathrm{top}(j_1,j_2)$ in the
system part decays with time, which is almost similar to the decay behavior
observed in the uniform case [compare with \figref{fig:sop:casei:part}].
However, very interestingly, the topological order parameters for larger
region, which are initially zero, become finite after some time passes [follow
the solid and dotted lines in \figref{fig:sop:caseii:cm}]. We find that the
time for $S_\mathrm{top}(j_1,j_2)$ with $j_1\leq N_\mathrm{env}$ to be maximally increased is approximately
$(N_\mathrm{env}+1-j_1)/w$: For example, the time when $S_\mathrm{top}(1,N)$ becomes
maximal is around $N_\mathrm{env}/w$. These findings strongly suggest that the
topological order propagates into the non-topological region with a constant
velocity after the quench.

The mechanism for the propagation of the topological order can be understood in
terms of the time evolution of Majorana wave functions, as in the uniform
case. \Figref{fig:sop:caseii:cm}(a) shows the snapshot of Majorana wave
function at $t = t_8$ when $S_\mathrm{top}(j_1,j_2)$ for $j_1 = 5$ is maximal with
$S_{\mathrm{top},x}(j_1,j_2) \approx S_\mathrm{top}(j_1,j_2)$ and $S_{\mathrm{top},y}(j_1,j_2) \approx 0$ as marked in \figref{fig:sop:caseii}(b). The
Majorana fermions which are initially inside the system part and form the
topological pairing are split and propagate into either direction. At $t=t_8$,
the main part of the Majorana wave functions covers the whole
region of Majorana sites for $S_\mathrm{top}(j_1,j_2)$ with $j_1=5$ as marked by
red vertical lines in \figref{fig:sop:caseii:cm}(a). It clearly reveals that
the propagation of the topological order is connected to that of the Majorana
wave functions, which determines the propagation speed.
The blue circles and lines connecting them in the figure pick some of the
Majorana pairings which make finite contribution to $S_{\mathrm{top},x}$ and
$S_\mathrm{top}$. Moreover, since the Majorana wave function has been spatially
split, the pairing of Majorana fermions contributing to the topological order
is now nonlocal: The Majorana fermions are bound not only to the nearest
neighboring one but also to that moving in the opposite direction with
increasing distance between them (see the connections with dotted lines). Note
that this nonlocal pairing leads to the vanishing of $S_\mathrm{top}(j'_1,j'_2)$
for almost all $j'_1 > j_1$: While the topological order arises in the region
$(j_1,j_2)$ for $j_1 = 5$ at $t = t_8$, it does not for smaller region. The
reason is that the pairing of Majorana fermions inside the smaller region is
not closed by themselves due to the nonlocal pairing. Some of Majorana fermions
are bound to those outside the region. It is quite different from the uniform
case in which the topological order is revived in all scales. On the other
hand, we would like to point out that even when $S_\mathrm{top}(j_1,j_2)$ is
maximal its value is rather small: For example, $S_\mathrm{top}(j_1,j_2) \approx
0.2$ for $j_1=5$. It is because the non-topological pairing of Majorana
fermions starting from the environment part is also propagating into that
region $(j_1,j_2)$.

The topological order eventually covers the whole wire,
which occurs at $t = t_9$ as seen in \figref{fig:sop:caseii}. At this time, the
Majorana fermions forming the topological pairing reach their outermost
boundaries which are $k=2$ and $2N-1$ for $S_{\mathrm{top},x}$ [see the red
vertical lines in \figref{fig:sop:caseii:cm}(b)]. So, although not perfect,
the Majorana pairings become similar to those in \figref{fig:Majoranachain}(b)
with additional nonlocal pairings.
After $t = t_9$, the Majorana fermions maintaining the topological pairing are
reflected at the boundaries and some of them going inside the wire so that the
region having the topological order shrinks (follow the solid lines in
\figref{fig:sop:caseii}). As the Majorana wave functions propagate, the
topological region should repeat expanding and shrinking. However, the
dispersion of the Majorana wave function causes topological order to gradually decay.
% so that no significant $S_\mathrm{top}(j_1,j_2)$ after
% the first shrinking.

\begin{figure}[!t]
  \centering
  \includegraphics[width=.6\textwidth]{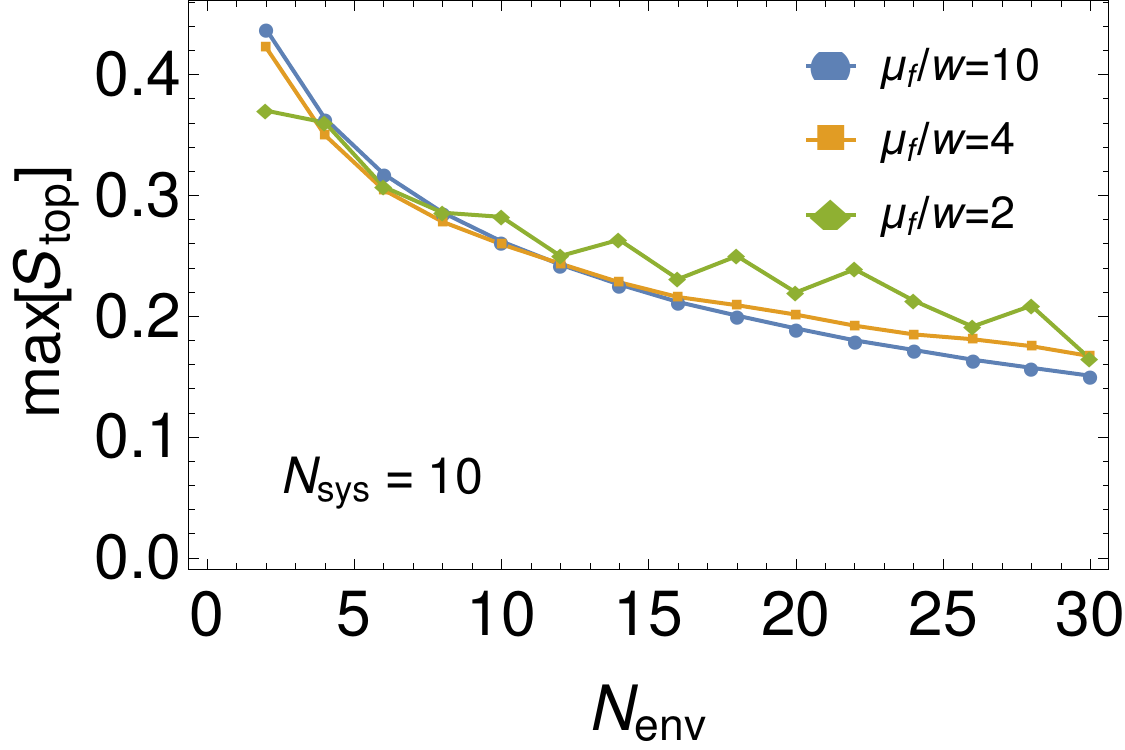}
  \caption{Maximum value of $S_\mathrm{top}$ when the topological order
    propagates over the entire wire at the first time as a function of
    $N_\mathrm{env}$ for different values of $\mu_f$. Here we use
    $N_\mathrm{sys} = 10$.}
  \label{fig:sop:caseii:maximum}
\end{figure}

While the topological region is determined mainly by the outermost boundaries of
Majorana wave function with topological paring, the inner boundaries of
Majorana fermions can also define the topological
region. \Figref{fig:sop:caseii:cm}(c) shows the Majorana wave function at $t =
t_{10}$ when $S_\mathrm{top}(j_1,j_2)$ for the system part ($j_1 = 11$) is
revived. While this revival is not due to the recombination of split Majorana
wave function as happens in \figref{fig:sop:casei:cm}(g), it can be also
explained in terms of the time evolution of Majorana wave functions.  First,
consider the initial edge Majorana fermions, $\gamma_{2\times11-1}$ and
$\gamma_{2\times 20}$ for the topological system part, connected by the red
dotted line in \figref{fig:sop:caseii:cm}. After the quench, they are split and
the two parts of each of them move outward and inward, respectively. So after
some time passes (around $t = t_{10}/2$) the inner parts of their split
Majorana wave functions cross each other. After that, they define the
boundaries of another topological region in which the topological pairing is
significant as shown in \figref{fig:sop:caseii:cm}(c). Such a revival
of the topological order happens along the dotted lines in
\figref{fig:sop:caseii}. Since the boundary of the topological region is moving
with time, this topological order also propagates along the wire.

Since the propagation of the Majorana wave function is accompanied with its
dispersion, the topological order when it covers the whole wire should be
weaker for longer wires. \Figref{fig:sop:caseii:maximum} clearly confirms
this expectation: $S_\mathrm{top}$ at its first maximum decreases monotonically
with increasing $N_\mathrm{env}$, except small additional oscillations observed
for $\mu_f$ close to $w$. It shows that as long as $\mu_f$ is large enough the
value of $\mu_f$ does not affect the propagation of the topological order as
observed in the uniform case [compare with \figref{fig:sop:casei}(c)].

\begin{figure}[t]
  \centering
  \includegraphics[width=.6\textwidth]{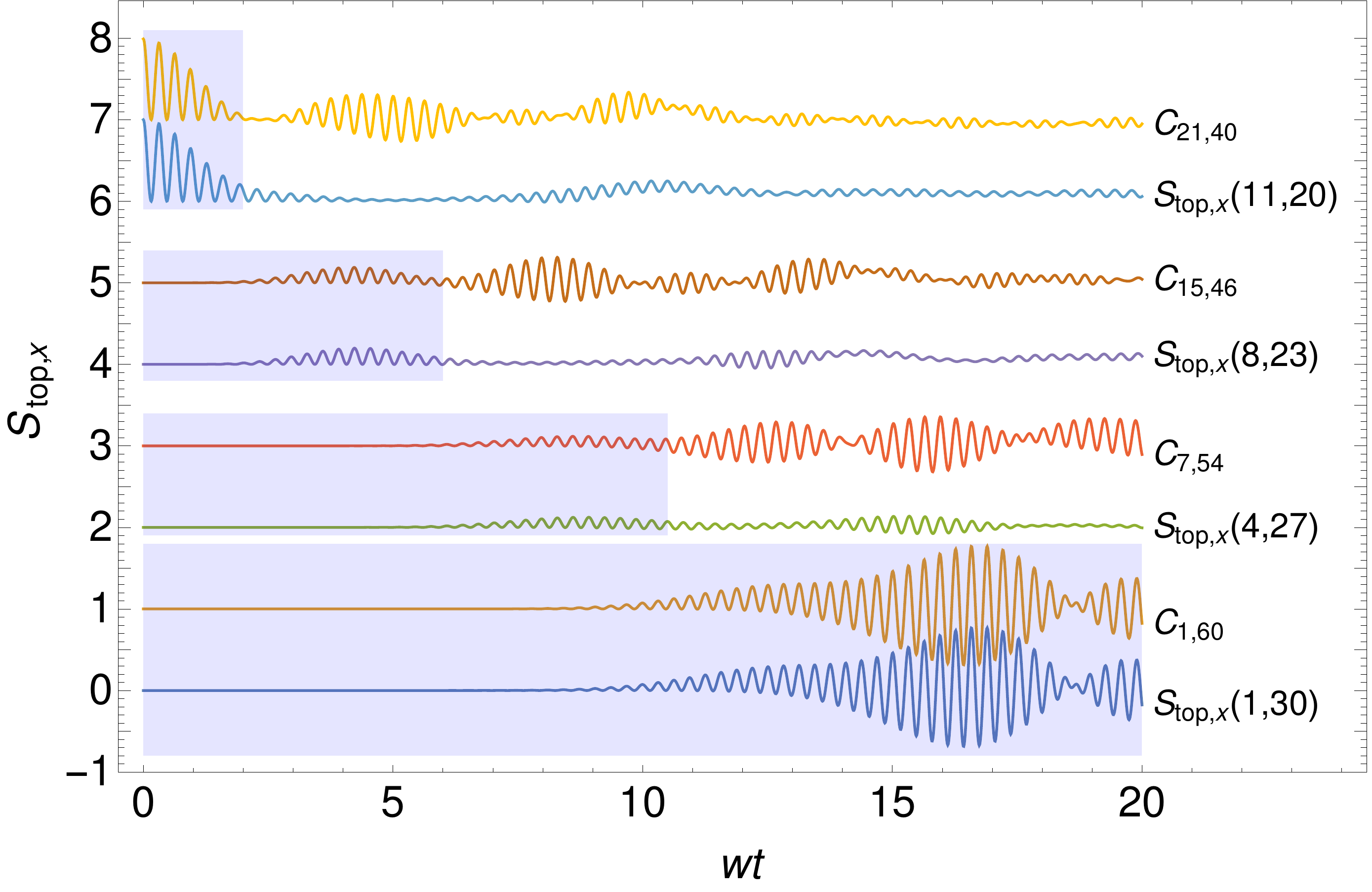}
  \caption{Comparison between the time evolution of $S_{\mathrm{top},x}(j_1,j_2)$
    and $C_{2j_1-1,2j_2}$ for the quench used in \figref{fig:sop:caseii}. The
    shaded boxes mark the time interval in which those two quantities are in
    perfect or good agreement with each other.}
  \label{fig:sop:caseii:relation}
\end{figure}

In the previous section, we observed interesting relations between the string
order parameters and the elements of the Majorana correlation matrix, such as
\eqnsref{eq:C:stopx} and (\ref{eq:C:stopy}), for the uniform wire.  Here we
demonstrate that similar relations hold in the non-uniform case as well. First,
as seen in \figref{fig:sop:caseii:relation}, the nonlocal topological order for
the whole system, $S_{\mathrm{top},x} = S_{\mathrm{top},x}(1,2N)$ is exactly
equal to the correlation $C_{1,2N}$ between Majorana fermions at the ends of
wire, which in fact comes from the exact identity, see
\eqnref{eq:C:stopx}. Then, one may question if a similar relation such as
$S_{\mathrm{top},x}(j_1,j_2) = C_{2j_1-1,2j_2}$ can hold for a part of the
system. As shown in \ref{sec:app:topologicalorder}, there is no such equation
identity. However, \figref{fig:sop:caseii:relation} demonstrates that the
relation can be valid at least for a finite time after the quench. The
temporary agreement between $S_{\mathrm{top},x}(j_1,j_2)$ and $C_{2j_1-1,2j_2}$
can be understood in terms of the general relation,
\eqnref{eq:stop:general}. Until the topological order spreads over the region
$(j_1,j_2)$, one can assume that the outer region remains strictly in the
non-topological phase so that the Majorana fermions in that region form the
non-topological pairings like in \figref{fig:Majoranachain}(a) and the string
order parameters $\Braket{\prod_{j=1}^{j_1-1} (-i\gamma_{2j-1} \gamma_{2j})}$
and $\Braket{\prod_{j=j_2-1}^{N} (-i\gamma_{2j-1} \gamma_{2j})}$ for the left-
and right-side environments are one. Therefore, together with
\eqnref{eq:stop:general}, we obtain
\begin{align}
  \label{eq:sop:approx}
  S_{\mathrm{top},x}(j_1,j_2)
  = \operatorname{Pf}[C_{\bar\varJ,\bar\varJ}]
  = C_{2j_1-1,2j_2}
\end{align}
since only $\gamma_{2j_1-1}$ and $\gamma_{2j_2}$ in $\bar\varJ$ are not forming
the non-topological pairing. The assumption leading to \eqnref{eq:sop:approx}
is broken once the topological order spreads beyond the region $(j_1,j_2)$.

\section{Experimental Detection of Topological Order}
\label{paper::sec:5}

While from the theoretical point of view the notion of the topological order is interesting and useful in understanding new quantum states of matter, its direct measurement is highly non-trivial even at the conceptual level since it is a nonlocal property. In this respect, the existence of the string order parameter is encouraging because (although nonlocal) it is suitable with current technology for experimental measurement \cite{Endres11a}.

\begin{figure}[t]
\centering
\includegraphics[width=70mm]{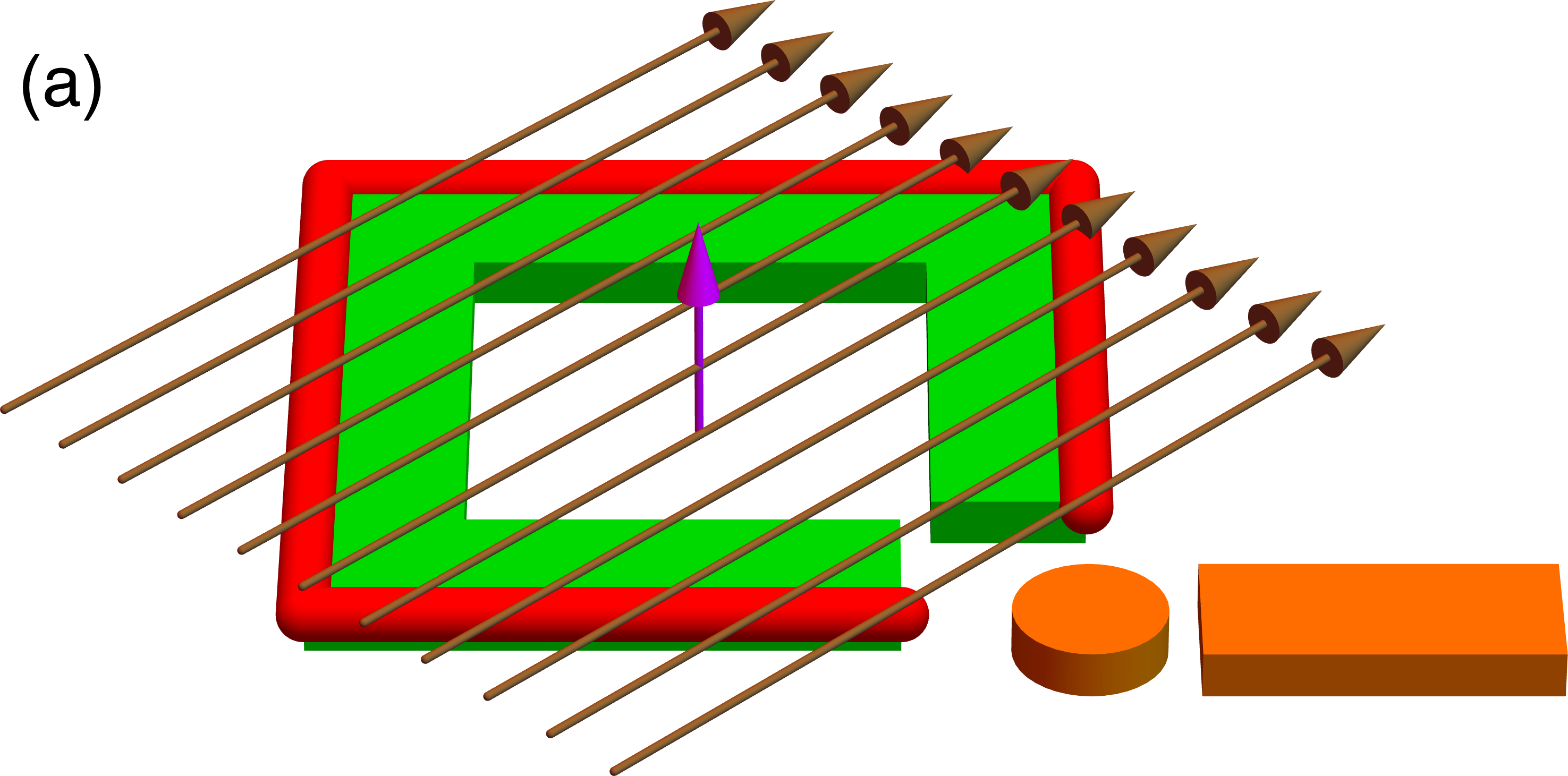}
\includegraphics[width=70mm]{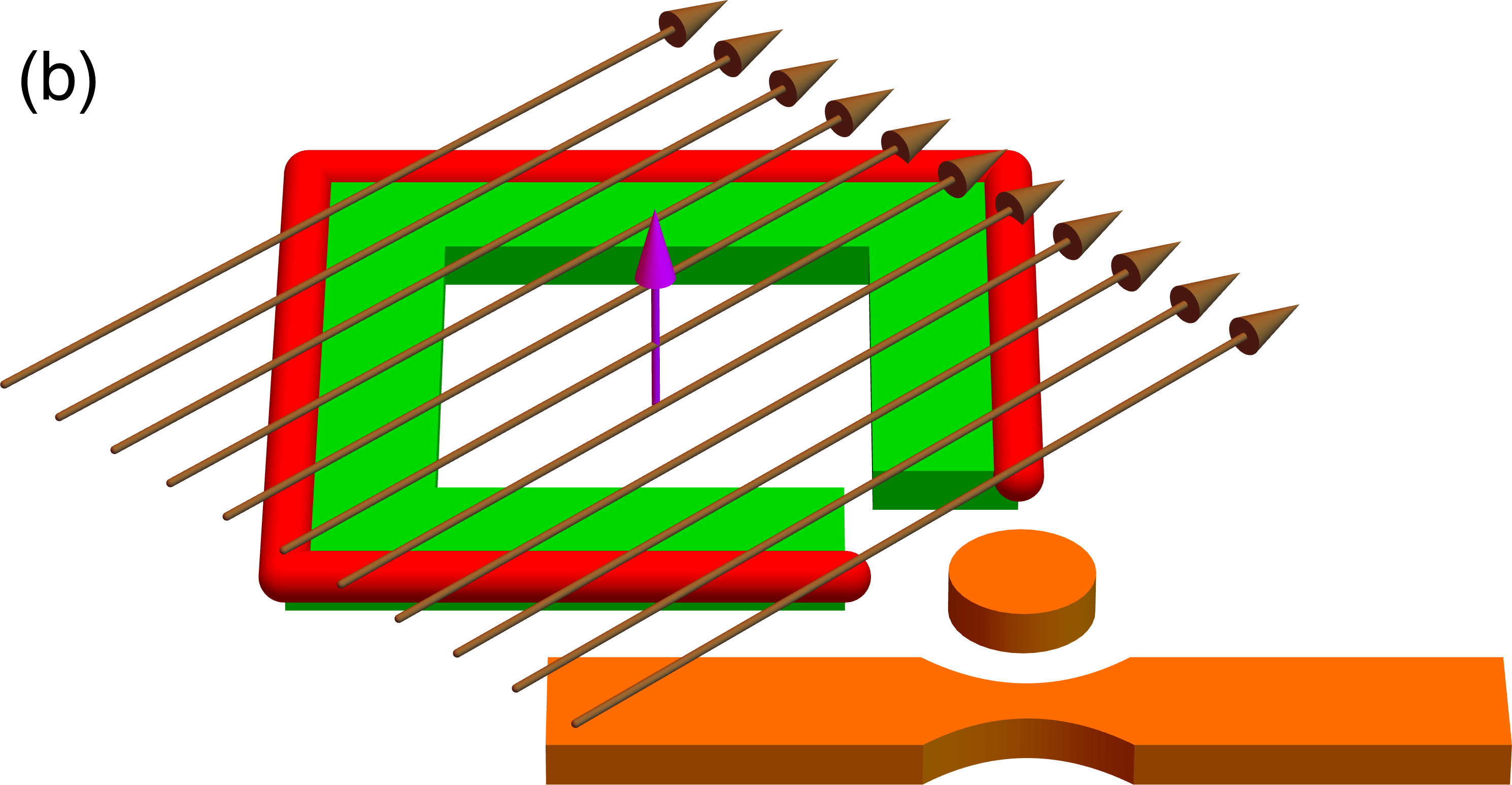}
\caption{(color online) Two schemes for measuring the occupancy of the
  hybridized states, $c_\pm = (c_1 \pm c_N)/\sqrt2$, between the two end-site
  electrons $c_1$ and $c_N$.  The wire (red) in the proximity to the
  conventional superconductor (green) and under the inplane magnetic field
  (black arrows) forms the topological superconductor. The two end-sites $c_1$
  and $c_N$ are coupled to a quantum dot (orange disc). Depending on the
  additional flux $\Phi$ (brown arrow at the center) through the loop, either
  $c_+$ ($\Phi/\Phi_0=0$) or $c_-$ ($\Phi/\Phi_0=\pi$) electron can tunnel to
  the quantum dot. (a) The excess electron on the quantum dot is probed by taking it out to an additional normal lead (orange bar) attached to the dot. (b) The excess electron is probed by continuously monitoring the current through the nearby quantum point contact (orange pinched bar).}
\label{fig:experiment}
\end{figure}

Here we propose a local and direct experimental way to probe the topological order. It is based on the relations between the string order parameters $S_{\mathrm{top},x/y}$ and the elements of the correlation matrix $C$ [see Eqs.~(\ref{eq:C:stopx}) and (\ref{eq:C:stopy})] discussed in Sections~\ref{sec:uniform} and \ref{sec:propagation}. First, note that the relations (\ref{eq:C:stopx}) and (\ref{eq:C:stopy}) allows us to rewrite the string order parameter $S_\mathrm{top}$ into
\begin{equation}
S_\mathrm{top}
 = C_{1,2N} - C_{2,2N-1}
= -2 \Braket{c_1^\dag c_N + c_N^\dag c_1}
= 2 (\Braket{c_-^\dag c_-} - \Braket{c_+^\dag c_+})
\end{equation}
where $c_\pm = (c_1\pm c_N)/\sqrt2$ are the fermion operators on the
hybridized fermionic states between two end sites. Therefore, the string order parameter can be measured by probing the occupancy of the hybridized state $c_\pm$ between the two end-site electrons $c_1$ and $c_N$.

Figure~\ref{fig:experiment} shows the schematics of two experimental setups to
probe the hybridized states $c_\pm$.
The quantum dot (the orange disc in Fig.~\ref{fig:experiment}) is tunnel coupled to the end sites $1$ and $N$ of the topological superconducting wire.
It is described by the Hamiltonian of the standard form
\begin{equation}
H_\text{probe} = -w_p\,d^\dag \left(e^{i\Phi/\Phi_0} c_1 + c_N\right)
+ h.c. \,,
\end{equation}
where $w_p$ is the coupling strength (assumed to be the same for $c_1$ and $c_N$ for simplicity), $d$ and $d^\dag$ are quantum dot electron operators, $\Phi$ is the additional flux through the loop (the brown arrow in Fig.~\ref{fig:experiment}) formed by the wire and the quantum dot, and $\Phi_0=h/e$ is the magnetic flux quantum. When $\Phi/\Phi_0=0$, the hybridized electron $c_+$ ($c_-$) can (cannot) tunnel onto the quantum dot because of the constructive (destructive) interference. For $\Phi/\Phi_0=\pi$, it is the other way around.
Therefore, by monitoring the excess electrons on the quantum dot, one can selectively detect the occupancies $\braket{c_\pm^\dag c_\pm}$.

To monitor the excess charge on the quantum dot, there are several schemes available in the present state-of-the-art technology. Here we consider two different schemes.
The first scheme [Fig.~\ref{fig:experiment} (a)] is, in spirit, the same as the quantum optical method of detecting the occupancy of excited atomic levels by observing the emitted photons~\cite{Scully97a} as no photon is emitted by a ground-state atom.
Operationally, it is similar to the on-demand single-electron emitter~\cite{Feve07a,Bocquillon13a,Gabelli06a}:
An excess electron on the dot is taken out to an additional normal lead $L$ attached to the dot by temporally manipulating the dot energy level $\epsilon_d(t)$. The dot-lead coupling is characterized by the level-broadening parameter $\Gamma_L$.
Throughout the measurement procedure, a bias voltage is applied so that $\mu_L<\mu$, where $\mu_L$ and $\mu$ are the chemical potentials of the normal lead and the system, respectively.
Initially, $\epsilon_d$ is set higher than $\mu$ so that no electron can tunnel into the quantum dot.
At the desired moment of measurement, $\epsilon_d$ is lowered and tuned between $\mu_L$ and $\mu$ ($\mu_L<\epsilon_d<\mu$). The electron (if any) occupying the hybridized state $c_\pm$ will tunnel into the quantum dot and then to the normal lead. After a certain period $\tau_p$ of time, $\epsilon_d$ is pushed back higher than $\mu$. This procedure is repeated with interval $\tau_r$ ($1/\tau_r$ is typically in the GHz range~\cite{Feve07a,Bocquillon13a,Gabelli06a}) many times enough for the electric current resolution.
When the system is in the topological state, the Majorana states at the ends of the wire are expected to be true bound states, the current is estimated to be
\begin{equation}
I_\pm \approx
\frac{4w_p^2\Gamma_L\tau_p}{\epsilon_d^2+4w_p^2}
\frac{e}{\tau_r}
\quad (w_p\ll\Gamma_L)
\end{equation}
if the hybridized state $c_\pm$ is occupied; the current vanishes otherwise.

In the second scheme [Fig.~\ref{fig:experiment} (b)], the excess electron on the quantum dot is probed by continuously monitoring the current through a nearby quantum point contact~\cite{Gurvitz97a,Elattari00a,Buks98a,Cassidy07a}.
It is possible because the dot electron casts an additional Coulomb potential and reduces the transmission probability from $T_\text{QPC}$ to $T_\text{QPC}'$ (hence the current from $I_\text{QPC}$ to $I_\text{QPC}'$) through the quantum point contact. The change in the current is given by
\begin{equation}
I_\text{QPC}-I_\text{QPC}' =
\frac{e^2}{h}(T_\text{QPC}-T_\text{QPC}')V_\text{QPC} \,,
\end{equation}
where $V_\text{QPC}$ is the voltage bias applied across the quantum point contact. Note that the continuous monitoring affects the tunneling between the hybridized state $c_\pm$ and the dot. For example, when the system is in the topological state, the coherent oscillation
of frequency $\sqrt{\epsilon_d^2+4w_p^2}$
between $c_\pm$ and $d$ is subject to the dephasing of rate
\begin{math}
\Gamma_\phi\equiv\left(\sqrt{T_\text{QPC}}-\sqrt{T_\text{QPC}}'\right)^2
\end{math} and
the gradual relaxation of rate
\begin{equation}
\Gamma_\text{mix} = \frac{4 w_p^2\Gamma_\phi}{\epsilon_d^2+\Gamma_\phi^2}
\end{equation}
if the hybridized state $c_\pm$ is initially occupied; the oscillation is missing otherwise.
The quantum point contact is often replaced with a single-electron transistor~\cite{Schoelkopf98a,Aasime01a,Lu03a}.
In either case, the change in charge on the quantum dot can be monitored as fast as the radio-frequency range.
More detailed analyses including the effects of back-action noise are referred to Ref.~\cite{Hackenbroich01a} and references therein.

\section{Conclusion}
\label{sec:conclusion}

We have considered a wire of topological superconductor and studied the
temporal evolution of the topological order upon a quantum quench across the
critical point in terms of the string order parameter.  Unlike topological
quantum numbers, which are commonly used to describe the equilibrium
topological order, the string order parameter is defined with respect to the
full dynamical wave function and naturally captures the dynamical evolution of
the topological order reflected in the wave function. It is considered to be
more suitable for experimental observations \cite{Endres11a}.
We have found that the topological orders vanish with a finite decaying time
and that the initial decaying behavior is universal in the sense that it does
not depend on the wire length and the final value of the chemical potential
(the quenching parameter).
The revival of the topological order in finite-size wires and the propagation
of the topological order into the region which was initially in the
non-topological state have been observed and explained in terms of the
propagation and dispersion of the Majorana wave functions.
Finally, we have found the exact relations between the string order
parameters and some local correlations, which are valid as long as the
fermion parity is well defined. Based on these relations we have proposed a
local probing method which allows to measure the topological order which is
supposed to be nonlocal.

\section*{Acknowledgments}

This work was supported by the the National Research
Foundation (Grant Nos. 2011-0030046 and 2015-003689) and
the Ministry of Education (through the BK21 Plus Project) of Korea.

\appendix

\section{Majorana Correlation Matrix and Topological Order}
\label{sec:app:topologicalorder}

In this appendix we prove that the string order parameter over a part of the
system is ultimately related to the correlation outside of that region and find
the exact relation between them.

The Majorana correlation matrix $C(t)$ defined in \eqnref{eq:C} has the
following properties as stated in the main text:
\begin{enumerate}
\item It is $2N\times2N$ real skew-symmetric matrix.
\item $\operatorname{Pf}[C] = 1$ due to the fixing of fermion parity.
\item It has eigenvalues of either $+i$ or $-i$. In other words, $C = V D V^t$
  where $V$ is an orthogonal matrix and
  \begin{equation}
    D
    =
    \bigoplus_{n=1}^N
    \begin{bmatrix}
      0 & 1
      \\
      -1 & 0
    \end{bmatrix}.
  \end{equation}
\end{enumerate}
From the property (iii), one can derive
\begin{equation}
  C^{-1}
  = V D^{-1} V^t
  = V (-D) V^t
  = - C.
\end{equation}

Now we recall the generalized Cramer's rule \cite{Gong2002}: For a nonsingular
$n\times n$ matrix $A$ and $n\times m$ matrices $X$ and $B$ satisfying $AX =
B$,
\begin{equation}
  \label{eq:cramersrule}
  \det X_{\varI,\varJ} = \frac{\det A_B(\varI,\varJ)}{\det A},
\end{equation}
where $\varI = \{i_1,i_2,\cdots,i_k\}$ and $\varJ = \{j_1,j_2,\cdots,j_k\}$ are
ordered sets of indices ($1\le i_1\le \cdots \le i_k \le n$ and $1\le j_1\le
\cdots \le j_k \le m$), $X_{\varI,\varJ}$ is the $k\times k$ submatrix of $X$
with rows in $\varI$ and columns in $\varJ$, and $A_B(\varI,\varJ)$ is the
$n\times n$ matrix formed by replacing the $i_s^\mathrm{th}$ column of $A$ by the
$j_s^\mathrm{th}$ column of $B$ for all $s=1,\cdots,k$.

In \eqnref{eq:cramersrule}, we substitute $A = C$, $X = C^{-1}$, $B =
1_{2N\times2N}$, and $\varJ = \varI$. Then, the Cramer's rule gives rise to
\begin{equation}
  \det C^{-1}_{\varI,\varI}
  = \frac{\det C_{\bar\varI,\bar\varI}}{\det C}
\end{equation}
since $\det A_B(\varI,\varI) = \det C_{\bar\varI,\bar\varI}$, where $\bar\varI$
(also assumed to be ordered) is complementary to $\varI$. Applying the
properties (i-iii) of the Majorana correlation matrix $C$, we get
\begin{equation}
  \det C_{\varI,\varI} = \det C_{\bar\varI,\bar\varI}
  \quad\text{or}\quad
  \operatorname{Pf}[C_{\varI,\varI}]^2 = \operatorname{Pf}[C_{\bar\varI,\bar\varI}]^2
\end{equation}
since $C_{\varI,\varI}$ and $C_{\bar\varI,\bar\varI}$ are also
skew-symmetric. Explicit comparison between $\operatorname{Pf}[C_{\varI,\varI}]$ and $\operatorname{Pf}[C_{\bar\varI,\bar\varI}]$ leads to
\begin{equation}
  \label{eq:C:cramersrule}
  \operatorname{Pf}[C_{\varI,\varI}] = (-1)^{P(\varJ)} \operatorname{Pf}[C_{\bar\varI,\bar\varI}],
\end{equation}
where the $P(\varJ)$ is the number of permutations needed to transform the set
of indices $\{1,2,\dots,2N\}$ to
$\{\underbrace{i_1,i_2,\cdots,i_k}_\varJ,\underbrace{1,\dots,i_1-1,i_1+1,\cdots,2N}_{\bar\varJ}\}$. \Eqnref{eq:C:cramersrule}
is the Cramer's rule applied to Majorana correlation matrix.

Now, we apply \eqnref{eq:C:cramersrule} to the string order parameter. (1)
Suppose $\varI = \{1,2N\}$. Then, \eqnref{eq:C:cramersrule} leads to
\begin{equation}
  \label{eq:C:stopx:proof}
  C_{1,2N}(t)
  = \operatorname{Pf}[C_{\varI,\varI}]
  = \operatorname{Pf}[C_{\bar\varI,\bar\varI}]
  = S_{\mathrm{top},x}.
\end{equation}
Also, with $\varI = \{2,2N-1\}$, we obtain
\begin{equation}
  \label{eq:C:stopy:proof}
  C_{2,2N-1}(t)
  = \operatorname{Pf}[C_{\varI,\varI}]
  = \operatorname{Pf}[C_{\bar\varI,\bar\varI}]
  = -S_{\mathrm{top},y}.
\end{equation}
These are very interesting relations. Seemingly, the nonlocal string order
parameters are replaced by local correlations between edge Majorana
fermions. However, it does not mean that the topological order can be local. To
the contrary, these relations insist that the topological order is really
nonlocal, which will be clarified in the case (2) below. Note that these
relations are valid only when the fermion parity is fixed.

(2) Suppose that $\varJ = \{2j_1,2j_1+1,\cdots,2j_2-2,2j_2-1\}$. Then,
\begin{equation}
  \label{eq:stop:general}
  S_{\mathrm{top},x}(j_1,j_2)
  = \operatorname{Pf}[C_{\varJ,\varJ}]
  = \operatorname{Pf}[C_{\bar\varJ,\bar\varJ}].
\end{equation}
It states that the string order parameter over a part of the system (for
example, from site $j_1$ to site $j_2$) is ultimately related to the
correlation outside of the part over which the topological order is
examined. This in turn seems to reflect the fact that the topological order is
a global (rather than local) property.

\section*{References}

\bibliographystyle{iopart-num}
\bibliography{paper}

\providecommand{\newblock}{}
\begin{thebibliography}{10}
\expandafter\ifx\csname url\endcsname\relax
  \def\url#1{{\tt #1}}\fi
\expandafter\ifx\csname urlprefix\endcsname\relax\def\urlprefix{URL }\fi
\providecommand{\eprint}[2][]{\url{#2}}
% Bibliography created with iopart-num v2.1
% /biblio/bibtex/contrib/iopart-num

\bibitem{Landau37a}
Landau L~D 1937 {\em Zh. Eksp. Teor. Fiz.\/} {\bf 7} 19

\bibitem{Landau37b}
Landau L~D 1937 {\em Zh. Eksp. Teor. Fiz.\/} {\bf 7} 627

\bibitem{Landau80a}
Landau L~D and Lifshitz E~M 1980 {\em Statistical Physics (Part 1)\/} 3rd ed
  ({\em Landau Course of Theoretical Physics\/} vol~5) (New York: Pergamon
  Press)

\bibitem{Klitzing80a}
Klitzing K~v, Dorda G and Pepper M 1980 {\em Phys. Rev. Lett.\/} {\bf 45}
  494--497

\bibitem{Tsui82a}
Tsui D~C, Stormer H~L and Gossard A~C 1982 {\em Phys. Rev. Lett.\/} {\bf 48}
  1559--1562

\bibitem{Nayak08a}
Nayak C, Simon S~H, Stern A, Freedman M and Sarma S~D 2008 {\em Rev. Mod.
  Phys.\/} {\bf 80} 1083

\bibitem{Hasan10a}
Hasan M~Z and Kane C~L 2010 {\em Rev. Mod. Phys.\/} {\bf 82} 3045--3067

\bibitem{Laughlin81a}
Laughlin R~B 1981 {\em Phys. Rev. B\/} {\bf 23} 5632--5633

\bibitem{Laughlin83b}
Laughlin R~B 1983 {\em Phys. Rev. Lett.\/} {\bf 50} 1395

\bibitem{Kane05a}
Kane C~L and Mele E~J 2005 {\em Phys. Rev. Lett.\/} {\bf 95} 146802

\bibitem{Kane05b}
Kane C~L and Mele E~J 2005 {\em Phys. Rev. Lett.\/} {\bf 95} 226801

\bibitem{Qi11a}
Qi X~L and Zhang S~C 2011 {\em Rev. Mod. Phys.\/} {\bf 83}(4) 1057--1110

\bibitem{Stephen64a}
Stephen M and Suhl H 1964 {\em Phys. Rev. Lett.\/} {\bf 13} 797

\bibitem{Abrahams66a}
Abrahams E and Tsuneto T 1966 {\em Phys. Rev.\/} {\bf 152} 416--432

\bibitem{Hohenberg77a}
Hohenberg P~C and Halperin B~J 1977 {\em Rev. Mod. Phys.\/} {\bf 49} 435

\bibitem{Perfetto13a}
Perfetto E 2013 {\em Phys. Rev. Lett.\/} {\bf 110}(8) 087001

\bibitem{Rajak2014apr}
Rajak A and Dutta A 2014 {\em Physical Review E\/} {\bf 89} 1--8

\bibitem{Hegde15a}
Hegde S, Shivamoggi V, Vishveshwara S and Sen D 2015 {\em New Journal of
  Physics\/} {\bf 17} 053036

\bibitem{Vasseur14a}
Vasseur R, Dahlhaus J~P and Moore J~E 2014 {\em Phys. Rev. X\/} {\bf 4}(4)
  041007

\bibitem{DeGottardi11a}
DeGottardi W, Sen D and Vishveshwara S 2011 {\em New Journal of Physics\/} {\bf
  13} 065028

\bibitem{Kibble76a}
Kibble T~W~B 1976 {\em Journal of Physics A: Mathematical and General\/} {\bf
  9} 1387

\bibitem{Kibble80a}
Kibble T 1980 {\em Physics Reports\/} {\bf 67} 183 -- 199

\bibitem{Zurek85a}
Zurek W~H 1985 {\em Nature\/} {\bf 317} 505--508

\bibitem{Zurek96a}
Zurek W~H 1996 {\em Physics Reports\/} {\bf 276} 177 -- 221

\bibitem{Bermudez09a}
Bermudez A, Patan\`e D, Amico L and Martin-Delgado M~A 2009 {\em Phys. Rev.
  Lett.\/} {\bf 102}(13) 135702

\bibitem{Bermudez10a}
Bermudez A, Amico L and Martin-Delgado M~A 2010 {\em New Journal of Physics\/}
  {\bf 12} 055014

\bibitem{ChoiMS15b}
Lee M, Han S and Choi M~S 2015 {\em Phys. Rev. B\/} {\bf 92} 035117

\bibitem{Endres11a}
Endres M, Cheneau M, Fukuhara T, Weitenberg C, Schau{\ss} P, Gross C, Mazza L,
  Ba{\~n}uls M~C, Pollet L, Bloch I and Kuhr S 2011 {\em Science\/} {\bf 334}
  200--203

\bibitem{Haegeman12a}
Haegeman J, P\'erez-Garc\'{\i}a D, Cirac I and Schuch N 2012 {\em Phys. Rev.
  Lett.\/} {\bf 109} 050402

\bibitem{Pollmann12a}
Pollmann F and Turner A~M 2012 {\em Phys. Rev. B\/} {\bf 86} 125441

\bibitem{Bahri2014apr}
Bahri Y and Vishwanath A 2014 {\em Physical Review B\/} {\bf 89} 155135

\bibitem{Kitaev2001oct}
Kitaev A~Y 2001 {\em Physics-Uspekhi\/} {\bf 44} 131

\bibitem{Alicea10a}
Alicea J 2010 {\em Phys. Rev. B\/} {\bf 81} 125318

\bibitem{Alicea11a}
Alicea J, Oreg Y, Refael G, von Oppen F and Fisher M~P~A 2011 {\em Nat Phys\/}
  {\bf 7} 412--417

\bibitem{Alicea12a}
Alicea J 2012 {\em Rep. Prog. Phys.\/} {\bf 75} 076501

\bibitem{Pfeuty1970mar}
Pfeuty P 1970 {\em Annals of Physics\/} {\bf 57} 79--90

\bibitem{Barouch1971feb}
Barouch E and McCoy B 1971 {\em Physical Review A\/} {\bf 3} 786--804

\bibitem{Calabrese2012julb}
Calabrese P, Essler F~H~L and Fagotti M 2012 {\em Journal of Statistical
  Mechanics: Theory and Experiment\/} {\bf 2012} P07022

\bibitem{Scully97a}
Scully M~O and Zubairy M~S 1997 {\em Quantum Optics\/} (Cambridge: Cambridge
  University Press)

\bibitem{Feve07a}
Feve G, Mahe A, Berroir J~M, Kontos T, Placais B, Glattli D~C, Cavanna A,
  Etienne B and Jin Y 2007 {\em Science\/} {\bf 316} 1169--1172

\bibitem{Bocquillon13a}
Bocquillon E, Freulon V, Berroir J~M, Degiovanni P, Pla{\c{c}}ais B, Cavanna A,
  Jin Y and F{\`e}ve G 2013 {\em Science\/} {\bf 339} 1054--1057

\bibitem{Gabelli06a}
Gabelli J, F{\'e}ve G, Berroir J~M, Pla{\c{c}}ais B, Cavanna A, Etienne B, Jin
  Y and Glattli D~C 2006 {\em Science\/} {\bf 313} 499

\bibitem{Gurvitz97a}
Gurvitz S~A 1997 {\em Phys. Rev. B\/} {\bf 56} 15215

\bibitem{Elattari00a}
Elattari B and Gurvitz S~A 2000 {\em Phys. Rev. Lett.\/} {\bf 84} 2047

\bibitem{Elattari00b}
Elattari B and Gurvitz S~A 2000 {\em Phys. Rev. A\/} {\bf 62} 032102

\bibitem{Buks98a}
Buks E, Schuster R, Heiblum M, Mahalu D and Umansky V 1998 {\em Nature\/} {\bf
  391} 871

\bibitem{Cassidy07a}
Cassidy M~C, Dzurak A~S, Clark R~G, Petersson K~D, Farrer I, Ritchie D~A and
  Smith C~G 2007 {\em Applied Physics Letters\/} {\bf 91}

\bibitem{Schoelkopf98a}
Schoelkopf R~J, Wahlgren P, Kozhevnikov A~A, Delsing P and Prober D~E 1998 {\em
  Science\/} {\bf 280} 1238

\bibitem{Aasime01a}
Aassime A, Johansson G, Wendin G, Schoelkopf R~J and Delsing P 2001 {\em Phys.
  Rev. Lett.\/} {\bf 86} 3376

\bibitem{Lu03a}
Lu W, Ji Z, Pfeiffer L, West K~W and Rimberg A~J 2003 {\em Nature\/} {\bf 423}
  422--425

\bibitem{Hackenbroich01a}
Hackenbroich G 2001 {\em Phys. Rep.\/} {\bf 343} 463

\bibitem{Gong2002}
Gong Z, Aldeen M and Elsner L 2002 {\em Linear Algebra and its Applications\/}
  {\bf 340} 253--254

\end{thebibliography}

\end{document}